\documentclass[prb,aps,floatfix,amsmath,amssymb,superscriptaddress]{revtex4}

\usepackage{mathtools}
\usepackage{amsfonts}
\usepackage{amssymb}

\usepackage{graphicx}
\usepackage[all]{xy}

\usepackage{amsthm}

\newtheorem{theorem}{Theorem}[section]
\newtheorem{lemma}[theorem]{Lemma}
\newtheorem{prop}[theorem]{Proposition}

\newcommand{\rg}{{\cal R}}
\newcommand{\spo}{f}

\newcommand{\be}{\begin{equation}}
\newcommand{\ee}{\end{equation}}

\begin{document}
\title{How Quantum Are Non-Negative Wavefunctions?}
\author{M. B. Hastings}
\affiliation{Station Q, Microsoft Research, Santa Barbara, CA 93106-6105, USA}
\affiliation{Quantum Architectures and Computation Group, Microsoft Research, Redmond, WA 98052, USA}
\begin{abstract}
We consider wavefunctions which are non-negative in some tensor product basis.  We study what possible teleportation can occur in such wavefunctions, giving a complete answer in some cases (when one system is a qubit) and partial answers elsewhere.  We use this to show that a one-dimensional wavefunction which is non-negative and has zero correlation length can be written in a ``coherent Gibbs state" form, as explained later.  We conjecture that such holds in higher dimensions.  Additionally, some results are provided on possible teleportation in general wavefunctions, explaining how Schmidt coefficients before measurement limit the possible Schmidt coefficients after measurement, and on the absence of a ``generalized area law"\cite{genarealaw} even for Hamiltonians with no sign problem. 

One of the motivations for this work is an attempt to prove a
conjecture about ground state wavefunctions which have an ``intrinsic" sign problem that cannot be removed by any quantum circuit.  We show a
weaker version of this, showing that the sign problem is intrinsic for commuting Hamiltonians in the same phase as the double semion model under the technical assumption that TQO-2 holds\cite{tqo2}.
\end{abstract}
\maketitle

Positive and negative signs play an incredibly important role in quantum mechanics.  For example, the basic phenomenon of interference relies on a cancellation between amplitudes of opposite sign.  When we restrict to certain signs being positive, in many ways the problem seems to simplify.  As an example, quantum Monte Carlo simulations based on a stochastic
sampling of the path integral become a powerful tool
in the study of interacting many-body systems when the Hamiltonian has all off-diagonal elements negative (such a property has long been referred to as ``having no sign problem" in physics and more recently has been called ``stoquastic" in quantum information theory).  When we restrict to Hamiltonians with no sign-problem, it is believed that the computational difficulty of estimating the ground state energy becomes smaller.  While this problem is QMA-complete for arbitrary local Hamiltonians (see Ref.~\onlinecite{qma} for the exact statements of what Hamiltonians may be
considered), it is StoqMA-complete\cite{stoqma} for such Hamiltonians with a restriction to no negative signs, and it is believe that StoqMA is strictly smaller than QMA, although there is no proof.

This property of having no sign problem is, of course, a basis dependent statement.  If one takes a generic Hamiltonian with no sign problem and conjugates this Hamiltonian by a generic short-depth local quantum circuit, the result generically will have a sign problem.  Conversely, given a Hamiltonian with a sign problem, it is quite possible that by some local quantum circuit, one might be able to bring the Hamiltonian to a form with no sign problem, simplifying the simulation.
However, we may expect that for many Hamiltonians, there is no local transformation that will remove the sign problem.
Complexity-theoretic evidence in favor of this statement is that we expect that StoqMA and QMA are distinct.  In more detail, if it were possible for all local Hamiltonians (indeed, all local frustration free Hamiltonians) to find a local transformation that mapped the Hamiltonian to one with no sign problem, then we would have QMA=StoqMA, because we could then reduce any problem in QMA to a problem in StoqMA where the witness includes a specification of the local quantum circuit, and the verifier checks that the quantum circuit indeed maps the Hamiltonian
to one with no sign problem, and then does the verification of the witness for this new Hamiltonian with no sign problem.  If the circuit has constant depth, then the new Hamiltonian still acts on a constant number of qubits (one can also allow sufficiently slowly growing depth).
However, it would be desirable to have a better understanding and to prove whether such an obstruction is present or not; further, it would be interesting to see if such an obstruction could happen for a {\it gapped} Hamiltonian.

This last question is a beautiful question that as far as we know was first raised by Matthew Fisher\cite{mpafconj}: does there exist a gapped local spin Hamiltonian whose ground state cannot be mapped by a local quantum circuit to a state that is non-negative in the computational basis? (i.e., that it is
a sum of basis states in the computational basis that have positive coefficients)
If so, let us say that ``the ground state of that Hamiltonian has an intrinsic sign problem".  Or, more generally, if there are multiple degenerate ground states and then
a gap to the rest of the spectrum, is it possible that there is no quantum circuit that maps the ground state subspace to a space which has an orthonormal basis of non-negative wavefunctions?  In this case, let us say that ``the ground state subspace of that Hamiltonian has an intrinsic sign problem".
Michael Freedman has conjectured\cite{mfconj} that the sign problem indeed is intrinsic for the ground state (or ground state subspace)  of such simple Hamiltonians as the Levin-Wen Hamiltonian $H_{LW}$ for the double semion model (see Ref.~\onlinecite{LW} for this Hamiltonian).  

Here we should remark that there also are two weaker questions one might ask for a Hamiltonian $H$.
The weakest result would be that there is no local quantum circuit which brings the Hamiltonian $H$ to a form with no sign problem; in this case we will say that ``the Hamiltonian $H$ has an intrinsic sign problem".  A slightly stronger result would be that for all gapped local Hamiltonians $H'$ which have the same ground state subspace as $H$, the Hamiltonian $H'$ has an intrinsic sign problem; in this case, we will say that ``all Hamiltonians in the same phase as $H$ have an intrinsic sign problem" (here ``phase" refers to a particular quantum phase of matter).
Note that indeed if the ground state subspace of a Hamiltonian has an intrinsic sign problem, then all Hamiltonians in that phase have an intrinsic sign problem, and if all Hamiltonians in the same phase as $H$ have an intrinsic sign problem then $H$ itself has an intrinsic sign problem, so these indeed would be weaker results.
We show a slightly weaker form of this result in the next section.  The results in that section, \ref{isp}, are logically independent from later results in the
paper, and the reader may choose to read only one or the other.

Then, partly motivated by the attempt to show the strongest result about the wavefunction itself, we study questions of states in which all amplitudes are non-negative in the computational basis from the perspective of quantum information theory.  This leads to some unusual questions.  We begin with a question of teleportation: given a tripartite system $A,B,C$ with $A$ and $C$ completely decorrelated, what possible correlations can appear between $A$ and $C$ after measurement on $B$?  (This can alternatively be regarded as a question of entanglement swapping.)   Without the sign problem restriction, we can in fact create perfect entanglement between $A$ and $C$.  However, with the restriction that the wavefunction for this tripartite system has no sign problem, we find that the entanglement is limited.  We give a detailed analysis in the case that $A$ is a qubit, and find that as the dimension of $C$ grows, the possible entanglement between $A$ and $C$ after measurement increases but for any finite dimension of $C$ it is not possible to have maximal entanglement.  In the qudit case, we give a less quantitative argument using compactness and still succeed in showing that the entanglement is bounded.  Using these results for tripartite systems, we are then able to show that for a line of qudits, for a wavefunction which is non-negative in the computational basis and which has zero correlation length, there is a ``coherent Gibbs state" form for the wavefunction.  That is, in the computational basis the wavefunction is the Gibbs state of some approximately local Hamiltonian.  We discuss ways in which our results might possibly be improved to show such a form in two dimensions.  This would then hopefully be a step toward answering this question about intrinsic sign problem for the double semion state as it may be possible to classify topological order in such coherent Gibbs states.

Additionally we give some other results on possible entanglement after measurement in systems without a sign constraint.  Also, we briefly sketch the absence of a generalized area law (in the sense of Ref.~\onlinecite{genarealaw}) for Hamiltonians with no sign problem.  The motivation for showing that was that I believe that many problems in quantum many-body physics will simplify if the Hamiltonian has no sign problem so it is worth showing that they do not simplify to such an extent that the generalized area law holds.

\section{Intrinsic Sign Problem For Commuting Hamiltonians in the Same Phase as the Double Semion}
\label{isp}
We now give the proof that there is an intrinsic sign problem for all commuting Hamiltonians with property TQO-2\cite{tqo2} in the same phase as the double semion model.  
The proof works for any topology, including spherical.  It is natural to conjecture that a stronger result holds, namely that if there is a commuting Hamiltonian with property TQO-2 and no sign problem, then it is in the same phase as a discrete gauge theory.

We show that
 \begin{theorem}
\label{intsp}
Consider a system containing a disk $D$ of radius $L$ (the topology of the system is not important).  Then for any $R$ sufficiently small compared to $L$
there does not exist a Hamiltonian $H'$ 
such that
\begin{itemize}
\item[1.] $H'$ is a sum of commuting projectors, each supported on a disk of radius at most $R$,
\item[2.] $H'$ has property TQO-2 up to some length $L^*\geq L$ (see Ref.~\onlinecite{tqo2} and also defined below),
\item[3.] there is a local unitary quantum circuit $U$ with range at most $R$ which maps the ground state subspace of $H$ onto the ground state subspace of $H'$, where $H$ is the double semion model Hamiltonian, and
\item[4.] $H'$ has no sign problem.
\end{itemize}
\end{theorem}
Here we define the ``range" of a quantum circuit as follows: consider a circuit which involves $r$ rounds of pairwise disjoint unitary gates, such that in each round each gate is supported on a set of diameter at most $s$; then, the range is $rs$.  Note that for any operator $O$, conjugation by
a quantum circuit of range $R$ gives an operator supported on the set of points within distance $R$ of the support of $O$.

The proof is by contradiction.  We 
suppose such that an $H'$ and such a $U$ did exist, and then derive a contradiction from lemmas \ref{looppos},\ref{iszero},\ref{isnonzero}.
Lemma \ref{looppos} shows the existence of operators with ceertain properties assuming such an $H'$ exists.  One of these
properties is that the matrix elements of these operators (in the computational basis) are non-negative; in this case
we say that the operator is ``entrywise non-negative".  Lemma \ref{iszero} shows that
a certain expectation value of these operators vanishes by showing that  is proportional to the trace of the cube of the $S$ matrix which vanishes for the double
semion model.  Lemma \ref{isnonzero} uses the entrywise non-negativity to show that the same expectation value is non-vanishing, giving the desired contradiction.

The expectation value used is related to ideas Ref.~\onlinecite{dsneqtc}.  In particular, we use an idea of computing elements of the $S$ matrix by expectation values of what is called a ``twist product" of operators (explained below).  We generalize this twist product to multiple operators, but we continue to refer
to it as a twist product.

First, some notation.
Write $H'=\sum_X Q'_X$, where the sum is over sets $X$ with diameter at most $R$ and each $Q'_X$ is a projector supported on set $X$.  Let $P'_X=1-Q'_X$.
In general, for any set $Y$, we define
\be
P'(Y)=\prod_{X \subset Y} P'_X.
\ee
Similarly, write $H=\sum_X Q_X$ and let $P_X=1-Q_X$ and define $P(Y)$ similarly. 

We now define TQO-2 for a Hamiltonian $H'$ which is a sum of commuting projectors.
Such a Hamiltonian $H'$ has property TQO-2 up to some length scale $L^*$ if the following holds for all contractible sets $C$ with diameter less than $L^*$.
Let $P'_0$ project onto the ground state subspace of $H'$.  Let $C_1$ be the set of points within distance $R$ of $C$ (thus, if $X$ has diameter $R$ and $X$ intersects $C$, then $X \subset C_1$).
Then, the reduced density matrix of $P'(C_1)$ on $C$ equals the reduced density matrix of $P'_0$ on $C$, up to multiplication by a scalar (the multiplication by a scalar is necessary since we have not normalized the traces of $P'(C_1),P'_0$ to $1$).

We note that in Ref.~\onlinecite{tqo2}, TQO-2 was defined to be the weaker property that the kernel of the reduced density matrix of $\prod_{X, X \cap C \neq \emptyset} P'_X$ on $C$ agreed with the kernel of the reduced density matrix of $P'_0$ on $C$.
However, as shown in Ref.~\onlinecite{tqo22}, this weaker property (that the kernels agree) implies that the reduced density matrix of $P'(C_2)$ on $C$ agrees with the reduced density matrix of $P'_0$ on $C$ up to multiplication by a scalar, if $C_2$ is defined to be the set of points within distance $2R$
of $C$.
So, at the cost of increasing the diameter of $X$ from $R$ to $2R$, we can assume the stronger TQO-2 property; i.e, the results shown here assuming the TQO-2 property defined here will imply the same results for a theory with the weaker TQO-2 property up to a redefinition of $R$.

Note that all the elements of $P'_X$ are non-negative; if this holds for an operator, we say that the operator is ``entrywise non-negative".
Note also that the product of any two entrywise non-negative operators is entrywise non-negative.
We will say that some set is ``far" from from some other set if the distance $d$ between them is large compared to $R$.  That is, there will be some minimum bound on $d/R$; this minimum bound will depend upon some geometric details; that is, we will often want it to be the case that if one takes some operator and acts on it by the quantum circuit it remains inside some set, but since we will have several sets that need to be far from each other, the distances will add.
Similarly, we will say that a distance $d$ is ``large" if it is large compared to $R$, again with the minimum bound on $d/R$ depending on geometric details.
We will say something is ``near" or ``small" if it is not ``far" or ``large".

Let $W_i$ be the loop operators of the double semion theory corresponding to dragging an anyon $i$ around a given closed loop, with
$i$ ranging over anyon types in the theory (in this case, these are $1,b,\overline b,b \overline b$).  These operators are unitary operators
which commute with the Hamiltonian.
These operators
can be read off from Eq.~44 of Ref.~\onlinecite{LW}, with $W_1,W_2,W_3,W_4$ of that paper corresponding to $W_1,W_b,W_{\overline b},W_{b \overline b}$.
We will also define operators
\be
W'_i=U W_i U^\dagger.
\ee

We also need the following lemma which relies on TQO-2:
\begin{lemma}
\label{containslemma}
Suppose $H'$ exists obeying properties $1-3$ of theorem \ref{intsp}.
Consider any set $Y$.  Let $Y_2$ be the set of points within distance $2R$ of $Y$.  Assume $Y_2$ is contractible and has diameter smaller than $L^*$.
Then, $U^\dagger P'(Y_2) U \leq P(Y)$.  That is, every vector in the image of $U^\dagger P'(Y_2) U$ is in the image of
$P(Y)$.
The same statement also holds if both $P_X$ and $P'_X$ are interchanged everywhere and $U$ is interchanged with $U^\dagger$.
\begin{proof}
We use the TQO-2 assumption (the Levin-Wen Hamiltonian obeys TQO-2 and $H'$ is assumed to have TQO-2).
Let $Y_1$ be the the set of points within distance $R$ of $Y$.
By TQO-2, every state in the image of $P'(Y_2)$ has a reduced density matrix on $Y_1$ which agrees with the reduced density matrix of the ground state of $H'$ on that set.  
Note that $UP(Y)U^\dagger$ is supported on $Y_1$, and so the expectation value of $UP(Y)U^\dagger$ on any state in the image of $P'(Y_2)$ agrees with
the expectation value of $UP(Y)U^\dagger$ in a ground state of $H'$.  
Therefore, the expectation value of $P(Y)$ on any state in the image of $U^\dagger P'(Y_2) U$ agrees with the expectation value of $P(Y)$ on a ground state of $H$, and so is equal to $1$.
\end{proof}
\end{lemma}

\begin{figure}
\includegraphics{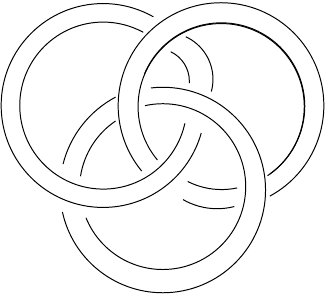}
\caption{A twist product of three operators on the three annuli.  The over and under crossings indicate the orderings.  This figure will be used below to compute the
trace of $S^3$, where $S$ is the $S$ matrix.}
\label{linkingcfig}
\end{figure}

\begin{figure}
\includegraphics{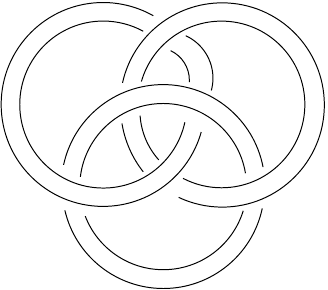}
\caption{A twist product of three operators on the three annuli.  The over and under crossings indicate the orderings.  This figure computes the
trace of $S^2 \overline S$, where $S$ is the $S$ matrix.  This figure is not used in the proof, but is merely used to illustrate that twist products different from Fig.~\ref{linkingcfig} can be used to compute other properties of the $S$ matrix.}
\label{linkingfig}
\end{figure}

We now construct the needed operators.  These operators will be supported on an annulus.
We do this construction for three different annuli, overlapping as in Fig.~\ref{linkingcfig} or in Fig.~\ref{linkingfig}; the crossings in those figures describes particular ``twist products" defined below.
\begin{lemma}
\label{looppos}
Consider a system containing a disk $D$ of radius $L$ (the topology of the system is not important).  Suppose $H'$ exists 
obeying properties $1-3$ of theorem \ref{intsp}.

Let $A$ be any annulus with large inner diameter and large distance between inner and outer boundary.
There exists an operator $L_A$ supported on $A$ such that
\be
\label{drags}
L_A=L^0_A\times \sum_i W'_i,
\ee
where $L^0_A$ is a non-vanishing operator supported on $A$, where $L_A$ commutes with $P'_X$ for all $X$, and
where the sum is over $i$ in 
$\{1,b,\overline b,b \overline b\}$.
where the loop defining $W'_i$ is chosen to be far from the edges of the annulus, winding around the annulus once.
Finally, $\sum_i W'_i$ commutes with $L^0_A$.

Further, if $H'$ has no sign problem, then $L_A$ is entrywise non-negative.

Further we show that $L_A$ is independent of the choice of loop so long as the loop is sufficiently far from the edges of the annulus.
\begin{proof}
Let $B$ be the set of points in $A$ which are distance greater than $2R$ from the boundaries of $A$.
Let $C$ be the set of points within distance $R$ of $A$.  So, $B$ is a thinner annulus and $C$ is a thicker annulus.
Let $T$ be the set of point in $A$ within distance $4R$ of the boundary of $A$.

We define $L_A$ as follows. 
Let $\rho'_{B}$ be the reduced density matrix of a ground state of $H'$ on $B$.  Note that since $A\subset D$, the reduced density matrix is independent of the choice of ground state if there is a degeneracy.
Then, let
\be
\label{LAdef}
L_A=P'(T)\rho'_{B} P'(T).
\ee
Since $P'_X$ and $\rho'_{B}$ are entrywise non-negative, $L_A$ is entrywise non-negative.

So, we now show Eq.~(\ref{drags}).
Let $\rho_{A}$ be the reduced density matrix of a ground state of $H$ on $A$.
Since $U$ has range $R$, the trace ${\rm tr}(O_B \rho'_B)$ for any operator $O_B$ supported in $B$ is equal to ${\rm tr}(U^\dagger O_B U \rho_A)$.
However, 
\be
\rho_{A}={\rm tr}_{C \setminus A}\Bigl({\rm const.} \times P(C)\times \sum_i W_i\Bigr),
\ee
as follows from known properties of the double semion model
(we omit a proof).
Note that the loop for $W_i$ can be taken to be any loop sufficiently far from the boundary.

Hence, since $U$ has range $R$ so that $U^\dagger O_B U\subset C$, the trace ${\rm tr}(O_B \rho'_B)$ is equal to ${\rm const.}\times
{\rm tr}\Bigl(U^\dagger O_B U P(C) (\sum_i W_i)\Bigr)$.
Hence,
\be
\rho'_B={\rm const.} \times  {\cal F}\Bigl(UP(C)(\sum_i W_i) U^\dagger\Bigr),
\ee
where ${\cal F}$ is the super-operator which traces out all sites not in $B$.
Write ${\cal F}$ as
\be
{\cal F}(O)=\int {\rm d}V \, V O V^\dagger,
\ee
where the integral is over unitaries $V$ supported on the complement of $B$, using Haar measure for the integration.
So,
\begin{eqnarray}
\rho'_B&=&{\rm const.} \times \int {\rm d}V \,  V UP(C) (\sum_i W_i) U^\dagger V^\dagger \\ \nonumber
&=& {\rm const.} \times  (\sum_i U W_i U^\dagger)  \int {\rm d}V \,  V UP(C)U^\dagger V^\dagger \\ \nonumber
& \equiv & (\sum_i U W_i U^\dagger) M_B,
\end{eqnarray}
where we use the fact that $[U  (\sum_i W_i)  U^\dagger,V]=0$ as follows from the assumption that $W_i$ is supported far from the boundaries of $A$, and
where $M_B$ is some operator supported on $B$ defined by the above equation.

So,
\be
\label{inmiddle}
L_A=   P'(T)M_B  P'(T)
(\sum_i U W_i U^\dagger).
\ee
Here we use the fact that $U W_i U^\dagger$ has disjoint support from $T$, so can be commuted through $P'(T)$ (we know that $W_i$ commutes with $P_X$
but we do not know that $U W_i U^\dagger$ commutes with $P'_X$).

Define $L^0_A=P'(T)M_B P'(T)$.
This operator is clearly non-zero.  Further, it commutes with $\sum_i W'_i$ since $\sum_i W'_i$ commutes with $M_B$.

We finally need to show that $L_A$ commutes with all $P'_X$.
Clearly it commutes with all $P'_X$ for $X \subset T$.  Also, clearly it commutes with all $P'_X$ for $X \not \subset A$.
So, we 
consider its commutator with $P'_X$ for $X$ in $A$ but not in $T$.  To show that this commutator vanishes, we show that for any such $X$,
$M_B P'_X = M_B = P'_X M_B$.  We show $M_B P'_X = M_B$ (the direction $P'_X M_B= M_B$ is similar).
So, we must show that
\be
 \int {\rm d}V \,  V U P(C)U^\dagger V^\dagger P'_X=
 \int {\rm d}V \,  V U P(C)U^\dagger V^\dagger.
\ee
However, $P'_X$ commutes with $V$ since the supports are disjoint.
So, we want to show
\be
U P(C)U^\dagger P'_X=
 U P(C)U^\dagger.
\ee
This however follows from lemma \ref{containslemma}.
\end{proof}
\end{lemma}

We now generalize the twist product of Ref.~\onlinecite{dsneqtc}.  In general, consider any set of operators $O_1,O_2,O_3,...,O_k$.  Consider any sets of sites $R_1,R_2,...,R_n$ with $R_a \cap R_b=\emptyset$ if $a \neq b$ and such that every $O_i$ is supported on $\cup_a R_a$.  Choose $n$ permutations, $\pi_j$, for $j=1,...,n$, each being a permutation on $k$ elements.  From this, define the ``twist product of $O_1,...,O_k$ with orderings $\pi_1,...,\pi_n$ on regions $R_1,...,R_n$"
as follows.
For each operator $O_i$, write $O_i=\sum_{\alpha} \prod_a O_{i,a}^\alpha$, where the sum is over sum index $\alpha$ and the product is over $a=1,...,n$ with $O_{i,a}^\alpha$ supported on $R_a$.
Then, the twist product is defined to be
$$\sum_{\alpha(1),\alpha(2),...,\alpha(k)} \prod_a \Bigl( O_{\pi_a(1),a}^{\alpha(\pi_a(1))}
 O_{\pi_a(2),a}^{\alpha(\pi_a(2))}
 \ldots 
O_{\pi_a(k),a}^{\alpha(\pi_a(k))}
\Bigr).$$
That is, we take the product of operators $O_i$, however on each region $a$ the operators are ordered by the permutation $\pi_a$.  Here, $\pi_a(b)$ denotes permutation $\pi_a$ applied to $b$.

This definition is independent of the choice of decomposition $O_i=\sum_{\alpha} \prod_a O_{i,a}^\alpha$.  One way to see this is to consider two different decompositions of a given operator.
To show that they give the same result, it suffices to consider their difference which gives some decomposition of the zero operator.
So, we need to show that if $O_i=0$ for some $i$, then the twist product vanishes.
For each choice of $\alpha(1),\alpha(2),...,\alpha(k)$ in the sum, the twist product is a product of $k$ different operators, these operators being
$ \prod_a  O_{\pi_a(j),a}^{\alpha(\pi_a(j))}$ for $j=1,...,k$.  Let us ``pad" this product with $k$ extra identity operators in front and back, writing the
twist product as a product of $3k$ operators: first, $k$ factors of $I$, then $ \prod_a  O_{\pi_a(j),a}^{\alpha(\pi_a(j))}$ for $j=1,...,k$, and finally
$k$ more factors of $I$.  We say that there are $3k$ ``time slices" and in each region on each time slice there appears an operator either $I$ or 
$O_{\pi_a(j),a}^{\alpha(\pi_a(j))}$.  Then, in any given region, we can change from having $k$ factors of $I$, followed by $O_{\pi_a(j),a}^{\alpha(\pi_a(j))}$
for $j=1,..,k$, and then $k$ more factors of $I$, to having $k_1$ factors of $I$, followed by $O_{\pi_a(j),a}^{\alpha(\pi_a(j))}$
for $j=1,..,k$, and then $k_2$ more factors of $I$ with $k_1+k_2=2k$.  In this way, by changing how many factors of $I$ appear before and after, we can
shift it so that in some time slice we have $O_{i,a}^{\alpha(i)}$ for all $i$.  Then, summing over $\alpha$, in this time slice we have $O_i=0$ giving zero for the twist product.

\begin{figure}
\includegraphics[width=1.5in]{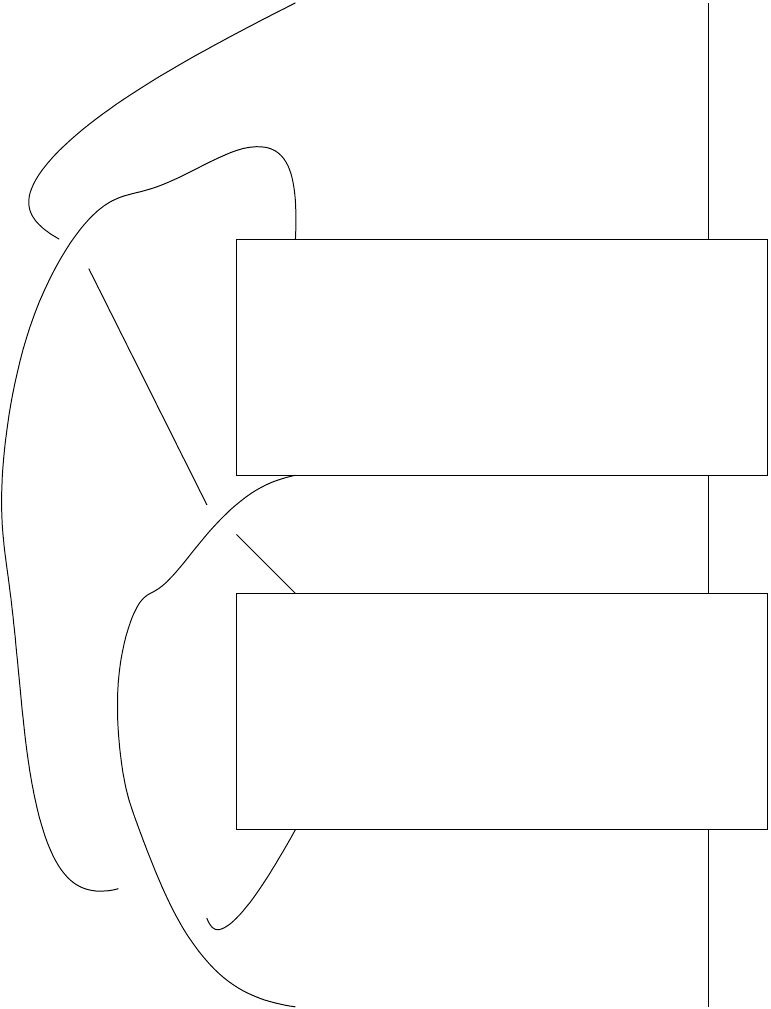}
\caption{Illustration of twist product written as tensor network.}
\label{tensortwistfig}
\end{figure}

It is also worth noting that the twist product can be written in a matter that does not use decompositions.  See for example Fig.~\ref{tensortwistfig}, which shows the twist product of two operators written as a tensor network.  Each square box represents an operator.  The lines incoming at the bottom of the box representing the state incoming to the operator (i.e., the column index of the operator when it is written as a matrix) and the lines outgoing from the top represent the state outgoing from the operator (i.e., the row index).  There are two lines as we have decomposed the Hilbert space as a tensor product of two Hilbert spaces, one on $R_1$ and one on $R_2$; the  left line represents one region and the right line represents another.

To denote the orderings in a twist product, it is useful to use pictures, as in Ref.~\onlinecite{dsneqtc}.  For twist products with three or more operators,
we can still use pictures.  A figure such as Fig.~\ref{linkingcfig} or Fig.~\ref{linkingfig} is used to indicate a twist product of three operators.  Each annulus indicates an operator
supported on that annulus.  The over and under crossings are used to indicate orderings on regions where two operators overlap; in more general products
where one might have three or more operators supported on a given site, one would again denote this with under and over crossings.  So, on any site which in only one annulus, the ordering of the operators does not matter since the other operators act as the identity on that site.  On any site in two annuli, we order the operators so that the operator on the undercrossing acts first; i.e., the ordering of the operators progresses vertically.

One useful property of the decomposition of operators used in the twist product is that
\begin{lemma}
\label{commlemma}
Suppose an operator $O$ supported on $R_1 \cup \ldots \cup R_n$ commutes with every term $P'_X$ in the Hamiltonian.  Then
$O$ can be decomposed as $\sum_\alpha \prod_a O_a^\alpha$ such that
each $O_a^\alpha$ commutes with every $P'_X$ except possibly for
those $X$ that have non-vanishing intersection with both $R_a$ and some $R_b$ for $b\neq a$.
\begin{proof}
Suppose first that there are only two regions $1,2$.
We decompose $O$ using a singular value decomposition
as $O=\sum_\alpha \lambda_a \prod_a O_a^\alpha$ where the $\lambda_a$ are scalars (using a Hilbert-Schmidt inner product to define the decomposition so that ${\rm tr}((O_1^\alpha)^\dagger O_1^\beta)={\rm tr}((O_2^\alpha)^\dagger O_2^\beta)=\delta_{\alpha\beta}$ where $\delta$ is the Kronecker delta).  In this way, every operator $O_1^\alpha$ is equal to ${\rm tr}_{R_2}(O O_2^\alpha)$ up to a scalar and similarly every $O_2^\alpha$ is
equal to ${\rm tr}_{R_1}(O O_1^\alpha)$.

If there are multiple regions, we use an iterated singular value decomposition: if there are regions $R_1,...,R_n$ then we first do a singular value decomposition of $O$ between $R_1$ and $R_2 \cup R_3 \cup \ldots \cup R_n$.  We then decompose every operator
on $R_2 \cup R_3 \cup \ldots. \cup R_n$ using a further singular value decomposition as a sum of a product of an operator on $R_2$ and an operator on $R_3 \cup \ldots \cup R_n$.  We then absorb the scalars (such as $\lambda_a$) that result from this decomposition into a redefinition of the operators in the singular value decomposition.

In this way, this gives a decomposition of $O$ as $O=\sum_{\alpha} \prod_a O_a^\alpha$ such that for every $O_a^\alpha$, there is an operator
$W_a^\alpha$ supported on ${\rm supp}(O)\setminus R_a$ such that
\be
\label{shows}
O_a^\alpha={\rm tr}_{{\rm supp}(O)\setminus R_a}(O W_a^\alpha),
\ee
where ${\rm supp}(O)$ denotes the support of an operator $O$.

If $X$ has vanishing intersection with $R_a$, then clearly $[O_a^\alpha,P'_X]=0$.
On the other hand, if $X$ has vanishing intersection with $R_b$ for all $b\neq a$, then
Eq.~(\ref{shows}) shows that $[O_a^\alpha,P'_X]=0$.
\end{proof}
\end{lemma}

\begin{figure}
\includegraphics{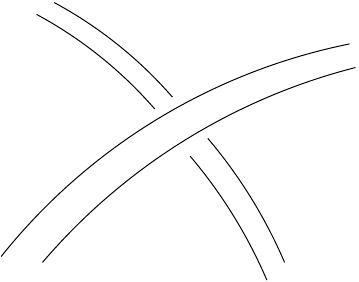}
\caption{A single crossing from Fig.~\ref{linkingcfig}.}
\label{onecrossingfig}
\end{figure}

\begin{lemma}
\label{iszero}
Consider a system containing a disk $D$ of radius $L$ (the topology of the system is not important).  Suppose $H'$ exists 
obeying properties $1-3$ of theorem \ref{intsp}.
Consider operators $L_{A_1},L_{A_2},L_{A_3}$ constructed as in lemma \ref{looppos}.
Then, the expectation value of the twist product in Fig.~\ref{linkingcfig}
is equal to zero.
\begin{proof}
The proof is based on showing that the expectation value is proportional to the trace of the cube of the topological $S$ matrix of the double semion model,
which is equal to zero.  In this lemma we do not use the positivity properties at all.

We introduce some notation, writing $W'_{i,1},W'_{i,2},W'_{i,3}$ to denote the operators $W'_i$ appearing in the constructions of $L_{A_1},L_{A_2},L_{A_3}$, respectively.
Note that 
\be
\label{equiv}
L_{A_c}=(1/4) L_{A_c} \sum_i W'_{i,c},
\ee
for $c=1,2,3$.
So, we will instead compute the twist product of three operators $(1/4) L_{A_c} \sum_i W'_{i,c}$.

We now give a decomposition to compute the twist product.  Note that
if given operators $M,N$ with decompositions $M=\sum_\alpha \prod_a M_a^\alpha$ and $N=\sum_\alpha \prod_a N_a^\alpha$, then we have a decomposition of
$MN$ as
\be
\label{MNdecomp}
MN=\sum_{\alpha,\beta} \prod_a M_a^\alpha N_a^\beta.
\ee
Using this identity, we separately decompose $L_{A_a}$ and $\sum_i W'_{i,a}$.  
We choose each region to contain at most one crossing and to contain the support of at most two of the operators $L_{A_a}$.
Consider a single crossing as in Fig.~\ref{onecrossingfig}.  We say that this crossing contains two ``arcs", where an arc refers to some thickened line
which is part of an annulus.
Suppose $L_{A_1},L_{A_2}$ are the two operators in this crossing.
We choose a region $\rg$ to contain this crossing.  Let $L_{A_1,\rg}^\alpha$ and $L_{A_2,\rg}^\alpha$ be the terms in the decomposition on this
region.  From lemma \ref{commlemma}, we can choose these terms so that  $L_{A_1,\rg}^\alpha$ and $L_{A_2,\rg}^\alpha$  commute with all $P'_X$ except those near the four endpoints of the two arcs.
We also need some decomposition of $W'_{i,c}$.  We will separately compute $4^3=64$ twist products corresponding to twist products of the
operators $L_{A_1} W'_{i,1}$, $L_{A_2} W'_{j,2}$, and $L_{A_3} W'_{k,3}$ and then sum over $i,j,k$.

Note that $W'_{i,c}$ commutes with $U P_X U^\dagger$ for all $X$, so we can choose a decomposition so that every term in the decomposition commutes with $U P_X U^\dagger$ for all $X$ except those near the end of an arc.  Let $(W'_{i,c,\rg})^\alpha$ be the terms in this decomposition.

Suppose further that in this crossing that $A_1$ overcrosses and $A_2$ undercrosses.
Then, the ordering of the operators in the crossing will be
$$L_{A_1,\rg}^\alpha (W'_{i,1,\rg})^\beta L_{A_2,\rg}^\gamma (W'_{j,2,\rg})^\delta.$$  (Note that $i$ need not equal $j$ here as $i$ and $j$ are being summed over).

We now consider the action of  $L_{A_2,\rg}^\gamma (W'_{j,2,\rg})^\delta$ on the ket $|\psi'\rangle$ where $\psi'$ is a ground state of $H'$.  Note
that $\psi'=U\psi$ for some ground state $\psi$ of $H$.  So, we will instead consider the action of
$U^\dagger L_{A_2,\rg}^\gamma U$ and $U^\dagger (W'_{j,2,\rg})^\delta U$ on $\psi$.
The action of $U^\dagger (W'_{j,2,\rg})^\delta U$ on a ground state of $H$ leaves the expectation value of all $P_X$ equal to $1$, except possibly for those
$X$ near the ends of the arc.  In fact, we can decompose this operator $U^\dagger (W'_{j,2,\rg})^\delta U$
as
\be
\label{dcmp}
U^\dagger (W'_{j,2,\rg})^\delta U = E_j D_j(2,\rg),
\ee
where $E_j$ acts near the endpoints of the arcs and $D_j(2,\rg)$ drags an anyon of type $j$ along the path in the arc $A_2 \cap \rg$; we write the
$2,\rg$ in parenthesis as we will apply similar decompositions to other operators and so we will need notation to indicate along which arc the anyon is dragged.
Here the operator $D_m$ that drags an anyon of type $m$ along some open loop is obtained from the definition of the loop operators $W_m$ in Ref.~\onlinecite{LW} by considering an open loop instead of a closed loop.  This decomposition (\ref{dcmp}) can be shown as follows.  The result just uses the fact that $U$ is a local quantum circuit and that
the operator $W_i$ in Ref.~\onlinecite{LW} can be written as a product of commuting operators, each supported on a small set, that is, we can write $W_{j,2}=\prod_x O(x)$ for some operators $O(m)$ all supported on small sets with $[O(m),O(n)]=0$ (we briefly explain this product at the end of this paragraph).
So, $W'_{j,2}=U W_{j,2} U^\dagger=\prod_x (U O(x) U^\dagger)$.  Now, we decompose the operator $W'_{j,2}$ as a sum of products of operators supported on the different regions.  Some of the operators $U O(x) U^\dagger$ will be supported on $\rg$; let the product of these operators be some operator $O(\rg)$.  Some other operators $U O U^\dagger$ are
supported on the complement of $\rg$; let the product of these operators be some other operator $O(\overline\rg)$.  Finally, there are some operators
$U O(x) U^\dagger$ supported both on $\rg$ and its complement; the product of these operators can be decomposed as a sum of products $\sum_{\alpha} O^\alpha(\rg) O^\alpha(\overline \rg)$ acting on $\rg$ and its complement.  So, this gives us a decomposition of $U O(x) U^\dagger$ as $\sum_{\alpha} \Bigl( O^\alpha(\rg) O(\rg) \Bigr) \Bigl(O^\alpha(\overline \rg) O(\overline \rg) \Bigr)$.  Then,$$U^\dagger O^\alpha(\rg) O(\rg) U=\Bigl( U^\dagger O^\alpha(\rg) U\Bigr) \Bigl( U^\dagger O(\rg) U \Bigr).$$  The second term in this product is the product of $O(x)$ for $x$ such that $U O(x) U^\dagger$ is supported on $\rg$; being a product of the operators $O(x)$ it is the operator that drags an anyon along an open loop and is the term $D_j$ in Eq.~(\ref{dcmp}) while the first term is supported near the endpoints of the arcs and is the term $E_j$.
Finally, we show how to write the operator $W_i$ given in Ref.~\onlinecite{LW} as a product of commuting operators.  The operator is given a product of operators on edges, ``R-legs", and ``L-vertices" (to use the terminology of Ref.~\onlinecite{LW}).  This product can be written as a product of commuting operators on small sets as follows; consider for example the operator $W_2$ using the notation of Ref.~\onlinecite{LW}; this is the operator that drags an anyon $b$.  This operator is
$$
\prod_{\rm edges} \sigma^z_j \prod_{\rm R-legs} i^{\frac{1-\sigma^x_j}{2}} \prod_{\rm L-vertices} (-1)^{s_I},
$$
where $s_I$ is defined in the reference.  One can only have at most $6$ L-vertices consecutively in the loop (after $6$ L-vertices, the loop closes), so after $6$ L-vertices either the loop must close or there must be an R-leg (R and L refer to different directions in which the loop turns).  Call a sequence of consecutive L-vertices an ``L arc".  We can then write the product above as
a product
$$\prod_{\rm L \, arcs} \Bigl(\prod_{\rm L-vertices \in L \, arc} (-1)^{s_I} \prod_{\rm edges \in L \, arc} \sigma^z_j  \Bigr) \prod_{\rm R-legs}  i^{\frac{1-\sigma^x_j}{2}} 
\prod_{\rm edges \, between \, R-legs} \sigma^z_j,$$
where the product for each L arc is over $L$-vertices in the given arc and over edges which touch at least one $L$-vertex in the arc.  The product over edges between $R$-legs is over
edges which touch two $R$-legs.  Then, this gives $W_2$ as a product of commuting operators; $W_3$ is similar, while $W_1,W_4$ are much easier to decompose.

Now consider the action of  $L_{A_2,\rg}^\gamma$ on the resulting state.  This operator leaves all expectation values $P'_X$ unchanged except near the
ends of the arc.  Hence, by lemma \ref{containslemma}, the action of $U^\dagger L_{A_2,\rg}^\gamma U$ leaves unchanged all $P_X$
except near the ends of the arc. 
We now use a property of the double semion model, which we discuss further at the end of the section.  Consider any operator $O$ (such as $U^\dagger L_{A_2,\rg}^\gamma U$ in this case) which is supported on some contractible set ${\rm supp}(O)$ (in this case, the arc) and which commutes with all terms in the Hamiltonian except those intersecting some other set $e_1 \cup e_2$ where $e_1,e_2$ are both contractible sets (in this case, $e_1,e_2$ will be the endpoints of the arc).  Consider any state $\phi$ which is a ground state of all terms in the Hamiltonian near ${\rm supp}(O)$, except possibly those near $e_1 \cup e_2$ (for example, the state $E_j D_j(2,\rg)\psi$ is such a state).  Then, the state $O\phi$
is equal to $\sum_m F_m D_m(2,\rg) \phi$ for some operators $F_m,D_m$, where the sum is over anyon types, where $F_m$ is supported near $e_1 \cup e_2$ and where $D_m(2,\rg)$ drags
an anyon of type $m$ from $e_1$ to $e_2$.

Similarly, considering the action of $U^\dagger L_{A_1,\rg}^\alpha U$ and $U^\dagger (W'_{i,1,\rg})^\beta U$ on the bra $\langle \psi|$.
These can also be written as $\sum_l G_l D_l(1,\rg)$ and $H_i D_i(1,\rg)$, where $D_i$ drags an anyon of type $i$ along the arc containing $A_1 \cap \rg$ and
$G_l$ and $H_i$ are supported near the ends of the arc.
So, we can replace these terms in the decomposition in this region with the new product of terms
$\sum_l G_l D_l(1,\rg) H_i D_i(1,\rg) \sum_m F_k D_m(2,\rg) E_j D_j(2,\rg)$.  In fact, we can make such a replacement in all regions, since
once one acts with terms (such as $ \sum_m F_k D_m(2,\rg) E_j D_j(2,\rg)$) in some region on the ket, the state still is a ground state of all terms in all
other regions, except possibly those near the boundary of those regions.
Having done this replacement, we have a new expectation value to compute.  For the product of operators in this expectation value, we have
the identity
\be
\label{reorder}
\sum_l G_l D_l(1,\rg) H_i D_i(1,\rg) \sum_m F_m D_m(2,\rg) E_j D_j(2,\rg)
=
\pm \sum_m F_m D_m(2,\rg) E_j D_j(2,\rg)
\sum_l G_l D_l(1,\rg) H_i D_i(1,\rg),
\ee
where the sign depends upon the anyon types $i,j,k,l$ (one must commute the dragging operators through each other).

We write the original twist product as a sum over $4^6$ terms, corresponding to the choices of $i,j,k$ in $W'_{i,1}$, $W'_{j,2}$, and $W'_{k,3}$, and to
the choices of $l,m,n$ in the operators which drag anyons in the decompositions of $L_{A_1},L_{A_2},L_{A_3}$; one might worry that an operator such as
$L_{A_1}$ might drag an anyon of one type along one arc in the decomposition, but of another type along another arc.  However, this gives zero contribution to the expectation value, since otherwise the operator produces a net topological charge at the end of an arc.

Now, fix $l,m,n$ and sum over $i,j,k$.  Using Eq.~(\ref{reorder}), we re-order the terms in the twist product until the annuli are no longer linked with
each other.  This requires changing the ordering at three of the crossings.
Note that the reason we consider the action of the terms on the ground state and show that we can replace them with equivalent terms is that we
have limited knowledge about the form of the operators $L_A$; however, once we show that their action on the ground state can be represented
using the operators which drag anyons then we have an explicit form for those operators and can compute the commutators exactly.

We claim that the sum over $i,j,k$ vanishes for any $l,m,n$.  To see this, note that after re-ordering, we can combine all the terms in the decomposition of
$W_{i,a}$ back into the original operator $W_{i,a}$.  Hence, up to signs, what we get is a product of operators, with each operator in the product either acting near the end points of the arcs, or dragging anyons of types $l,m,n$ along the arc, or dragging an anyon of type $i,j,k$ around a loop. 
Let $i'$ be the anyon type given by fusing $i,l$.
Similarly, let $j',k'$ be the anyon types given by fusing $j,m$ or $k,n$, respectively.
For fixed $l,m,n$, the sum over $i,j,k$ causes the anyon type $i',j',k'$ in the fused operator to range over all four possibilities; that is, if we fix $l,m,n$ and sum over
over $i',j',k'$, this is the same as fixing $l,m,n$ and summing over $i,j,k$.

The expectation value of this product of operators is independent of $i,j,k$.  Consider any set $e$ which is the endpoint of some arc.
Note that after recombining the terms in the decomposition of $W_{i,a}$ back into $W_{i,a}$, we are left with only two possible operators which might
not commute with terms $P_X$ for $X$ near $e$.  These are terms from the two arcs ending near that endpoint, such as $\sum_m F_m D_m(2,\rg)$ and a similar term $F'_m D_m(2,\rg')$ from the arc which is the intersection of $A_2$ with some other region $R_2$.  The only other operator acting near the endpoints $e$ is an operator $W_{i,a}$ which commutes with all terms in $H$.  
This operator $W_{i,a}$ occurs either before or after both terms $\sum_m F_m D_m(2,\rg)$, $F'_m D_m(2,\rg')$ so the terms
$\sum_m F_m D_m(2,\rg)$, $F'_m D_m(2,\rg')$ appear between a bra and a ket which both have expectation value $+1$ for all $P_X$ near $e$.
So, we can replace any operator $O$ in $\sum_m F_m D_m(2,\rg)$ or $F'_m D_m(2,\rg')$ which is supported near some endpoint $e$ by the operator $(\prod_{X, X \cap {\rm supp}(O) \neq \emptyset} P_X) O (\prod_{X, X \cap {\rm supp}(O) \neq \emptyset} P_X)$ and obtain the
same expectation value.  The resulting operator, however, acts trivially on the ground state; then, since the $W_{i,a}$ also act trivially on the ground state,
the result is independent of $i,j,k$.

Hence, the only effect of $i,j,k$ is to give an overall sign due to the sign in Eq.~(\ref{reorder}).  
The sign of Eq.~(\ref{reorder}) for commuting $i,l$ through $j,m$ is the {\it same} as that for commuting $i'$ through $j'$, and similarly for the other signs.  So the net sign depends only on the choice of $i',j',k'$ and after summing over $i',j',k'$ it gives
the trace of $S^3$ which vanishes.
\end{proof}
\end{lemma}

Finally,
\begin{lemma}
\label{isnonzero}
Suppose the operators $L_{A_1},L_{A_2},L_{A_3}$ in the twist product in Fig.~\ref{linkingcfig}
are entrywise non-negative and commute with all $P'_X$ for some Hamiltonian $H'_X$ which obeys TQO-2 and has no sign problem.
Then, the twist product is positive.
\begin{proof}
Note that for an operator that is entrywise non-negative and that acts on some set of disjoint regions $R_a$, we can choose a decomposition of that operator as a sum of products of operators, with each operator in the product acting on only one region and
such that every operator in the product is entrywise non-negative.
The specific decomposition that we choose is as follows.  For each region $R_a$, let $\psi_a(i)$ be a basis of states in the computational basis.
Let $O_a(i,j)=|\psi_a(i)\rangle\langle \psi_a(j)|$.  Then, to decompose an operator $L_A$, we separately decompose each operator $\rho'_B$ or $P'_X$ appearing in Eq.~(\ref{LAdef}) as a sum of products
of operators $O_a(i,j)$, and we use these decompositions to give a decomposition of $L_A$ using Eq.~(\ref{MNdecomp}) as a product of operators.

We write a basis for the ground
state wavefunctions using states that have only non-negative coefficients in the computational basis.  We will compute the expectation value
for any one of these basis states; we call this state $\Psi_0$.  By the disk axiom (which holds for the double semion model and also for the ground
states of $H'$ since they are obtained by acting with a local quantum circuit on the ground states of the double semion model), once we show that this expectation value is positive, this will imply that the expectation value
is positive for all ground states.
The disk axiom is also known as TQO-1 in Ref.~\onlinecite{tqo2}, and implies that all ground states have the same reduced density matrix on any region of diameter smaller than some $L^*$.

The state $\Psi_0$ is a sum of basis states in the computational basis with non-negative coefficients, and we have written the twist product as a sum of products of operators $O_a(i,j)$ with non-negative coefficients.  So, to show that the result is positive, it suffices to pick any one basis state $\Psi$ in the computational basis and any particular choice of operators $O_a(i,j)$ appearing in the twist product and show that the expectation value of that choice of operators
in state $\Psi$ is nonzero.  
We pick $\Psi$ to be any arbitrary computational basis state that has nonzero amplitude in the given ground state wavefunction $\Psi_0$; let $p$ be the square of this amplitude.  So, the expectation value of the twist product in state $\Psi_0$ is greater than or equal to $p$ times the expectation value of
the twist product in the basis state $\Psi$, which in turn is lower bounded by $p$ times the expectation value in the basis state $\Psi$ of any particular term in the sum in this
decomposition of the twist product.

The terms in the decomposition that we choose are the diagonal terms, of the form $O_a(i,i)$; that is, we consider only the diagonal terms in
the decomposition of  each operator, either $P'_X$ or $\rho'_B$.  So, we must show that for each $P'_X$ or $\rho'_B$, that $\langle \Psi | P'_X | \Psi \rangle >0$ and $\langle \Psi | \rho'_B | \Psi \rangle>0$.
This follows for $P'_X$ because $\Psi$ has positive overlap with $\Psi_0$ which is an eigenstate of $P'_X$ with eigenvalue $+1$.
It follows for $\rho'_B$ since $\rho'_B$ is the reduced density matrix of $\Psi_0$ which is equal to the reduced density matrix of $p|\Psi\rangle\langle\Psi|$ plus non-negative terms arising from contributions of other computational basis states in the wavefunction $\Psi_0$.
\end{proof}
\end{lemma}

Finally, we explain the claim that in the double semion model, given any operator $O$ which is supported on some contractible set ${\rm supp}(O)$ and which commutes with all terms in the Hamiltonian except those intersecting some other set $e_1 \cup e_2$ where $e_1,e_2$ are both contractible sets, and given any state $\phi$ which is a ground state of all terms in the Hamiltonian near ${\rm supp}(O)$, except possibly those near $e_1 \cup e_2$, the state $O\phi$
is equal to $\sum_m E_m D_m \phi$ for some operators $E_m,D_m$, where the sum is over anyon types, where $E_m$ is supported near $e_1 \cup e_2$ and where $D_m$ drags
an anyon of type $m$ from $e_1$ to $e_2$.
Consider the space of ground states of the Hamiltonian which is a sum of $Q_X$ over all $X$ that
are near ${\rm supp}(O)$, except possibly those near $e_1 \cup e_2$.
Let us say that the ``far boundary" is the set of points far from ${\rm supp}(O)$, and let us say that the boundary is the union of the far boundary and $e_1,e_2$.
Consider any configuration of closed loops in the $Z$ basis, except that the loops may be open on $e_1$ or $e_2$ or far from ${\rm supp}(O)$, i.e., they may be open on the boundary.  Acting on such a configuration with $\prod_{X \, {\rm near} \, {\rm supp}(O), X\cap (e_1 \cup e_2)=\emptyset} P_X$
gives a ground state if it gives a nonzero state, and all ground states can be obtained in this fashion because the projectors $P_X$ will annihilate any state with an open loop.
Any two states with the same configuration on the boundary and with the configuration in the same relative $Z_2$ homology class (relative to boundary) give the same ground state. 
This is a known result for the toric code\cite{boundary}; the double semion model has the same ground states up to a sign given by $-1$ raised to a power equal to the number of surgery moves required to turn a loop configuration into an arbitrary reference configuration\cite{LW}; on a sphere, this power is simply equal to $-1$ to the number of closed loops while with boundaries it is more complicated (on some non-orientable geometries, some of the toric code ground states cannot be given a sign structure consistent with
the double semion Hamiltonian\cite{WW}; while this does not happen here, even if it did happen, it would not affect the claim that all ground states could be described in this way, as then the ground states of the double semion model would simply be given by taking some subset of these toric code states, i.e., those which can be given an appropriate sign structure).
So, this gives us a basis for the ground state subspace, specified by where ends of open loops are located on the boundary and by homology class, if any.

The usefulness of this basis is that we can now show that we can act arbitrarily in this subspace by dragging anyons as follows.
One can detect open loops at the boundary using the operator which measures in the $Z$ basis at the boundary; we can regard this operator as dragging a $\overline b b$ anyon around
a very small open loop containing a single bond.  Thus, one can detect the state at the boundary by dragging anyons near the boundary.  One can change which loops are open at the
boundary by dragging an anyon of type $b$ from one point on the boundary to another; at both endpoints, this has the effect of creating an open loop if there was none, and destroying an open loop if there was one.  Thus, we can act on the configuration on the boundary arbitrarily by dragging anyons, and these operators all preserve this ground state subspace.
We have assumed that ${\rm supp}(O)$ is contractible, but since we have also sets $e_1,e_2$ which might
be ``punctures" within ${\rm supp}(O)$, in which case in fact there might be nontrivial homology, for example if the set of points away from the boundary has the topology
of a twice punctured disk. 
For a given
choice of open loop ends on the boundary, one can change homology class by
dragging an anyon $b$ around a loop which; since ${\rm supp}(O)$ is contractible, such loops supported in ${\rm supp}(O)$ can all be contracted to loops near $e_1$ or $e_2$ or a sum of loops around both.  Similarly, one can deform any operator which detects homology supported in ${\rm supp}(O)$ to an operator dragging anyon $b \overline b$ around loops near $e_1,e_2$.
Hence,
the action of any operator in the ground state subspace can be written as a sum of operators, each of which is a product of operators dragging anyons of various types from one place on the boundary to another, or dragging a $b$ around an element of homology, or dragging a $b \overline b$ around loops near $e_1,e_2$.  
Finally, once we have written an operator as a sum of products of operators dragging anyons, we deform the paths along which the anyons are dragged until we have only (for an
operator supported on ${\rm supp}(O)$), anyons dragged near $e_1$ or $e_2$ or from $e_1$ to $e_2$.
Any operator supported on ${\rm supp}(O)$ cannot change which loops are open on the far boundary, so we can restrict to sums of operators which do not drag $b$ from one point on the boundary to another.  Further, any operator supported on ${\rm supp}(O)$ commutes with all operators supported on the far boundary, so they can be sensitive only to the parity (even or odd) of open loops on the far boundary; however, the parity of open loops on the far boundary corresponds to dragging $b \overline b$ around the far boundary which can be contracted to dragging $b \overline b$ around loops near $e_2,e_2$.  So, the operator supported on ${\rm supp}(O)$ can be written as a sum of operators dragging anyons of various types
from some point near $e_1$ or $e_2$ to some point near $e_1$ or $e_2$.  For each term in the sum, we can compute the change in topological charge near $e_1$: we compute the total
anyon charge dragged from $e_1$ to $e_2$; this is $1,b,\overline b$, or $b \overline b$.  Then, we can write that term in sum as an operator $D_m$ which drags an anyon with the given total anyon charge from $e_1$ to $e_2$ multiplied by an operator $E_m$ supported near $e_1 \cup e_2$.

\subsection{Remarks on Computing the $S$ Matrix From the Ground State Wavefunction}
One might wonder whether this proof above could be adapted to prove the stronger result that the ground state of the double semion model has an
intrinsic sign problem.  In the proof above, we used the operators $P'_X$ defined from $H'$.  Unfortunately, if we are just given some wavefunction $\Psi_0$
that is obtained by acting with a local quantum circuit on a double semion model ground state such that $\Psi_0$ is non-negative in the computational
basis, there does not seem to be any easy way to construct a commuting projector Hamiltonian with no sign problem such that $\Psi_0$ is a ground state
of that Hamiltonian.  Note that given a density matrix $\rho$ that is entrywise non-negative, the projector $Q$ onto the kernel of $\rho$ may have a sign problem.  That is, a positive wavefunction might not have a parent Hamiltonian\cite{pH} without a sign problem.

Another possibility would be to instead define the operators $L_A$ to simply equal $L_A=\rho'_B$, without the extra projectors as in Eq.~(\ref{LAdef}).
In this case, it is still possible to prove a lemma like lemma \ref{isnonzero}.  However, proving a lemma like lemma \ref{iszero} become more difficult.  The operators $\rho'_B$ do not commute with the Hamiltonian near the boundary and so they may create additional anyonic excitations near the boundary.
This complicates the calculation of the commutation relations of these operators, complicating the calculation of the twist product.  It is not clear whether or
not a result like lemma \ref{iszero} would still continue to hold (perhaps one can at least show that the expectation value is small).

However, this may still give a way of computing the trace of $S^3$ (or other properties of the $S$ matrix) from the ground state wavefunction.
Let $\rho'_{B_1},\rho'_{B_2},\rho'_{B_3}$ be reduced density matrices on three annuli as in Fig.~\ref{linkingcfig}.
While it may not be true in general, it is reasonable to hope that for physical systems that the expectation value of the twist product of these reduced density matrices  is close to the trace of $S^3$ multiplied by the expectation value of their untwisted product (i.e., the expectation value of $\rho'_{B_1}\rho'_{B_2}\rho'_{B_3}$.
This may be accessible using quantum Monte Carlo and may give an alternative method to that suggested in Ref.~\onlinecite{Sgswf} to extract the
$S$ matrix from the ground state wavefunction.

\section{Teleportation}
We next turn to questions of positive wavefunctions, without any assumptions about an underlying Hamiltonian.  As a preliminary, we consider teleportation in tripartite wavefunctions, before proceeding to larger systems.

Consider a tripartite system, with parts labelled $A,B,C$ and corresponding Hilbert spaces ${\cal H}_A, {\cal H}_B, {\cal H}_C$.  Consider a wavefunction $\psi$ with amplitudes $\psi(i,j,k)$ in a product basis $|i\rangle \otimes |j\rangle \otimes |k\rangle$ with the 
property that the wavefunction is non-negative in the computational basis, i.e. that
\be
\forall i,j,k \quad \psi(i,j,k)\geq 0.
\ee
For notational shorthand, we refer to a wavefunction which is non-negative in the computational basis simply as a ``non-negative wavefunction".
Assume further that the density matrix $\rho_{AC}$ on $AC$ factors as
\be
\label{fac}
\rho_{AC}=\rho_A \otimes \rho_C.
\ee

We now consider the question of ``teleportation" in such a wavefunction: after a projective measurement on $B$ {\it in the same basis of states} $|j\rangle$, what entanglement can be produced between $A,C$?
Note that without the non-negativity constraint, it possible for $A,C$ to be maximally entangled: consider a system of four qubits, with $A$ being the first qubit, $C$ being the fourth qubit, and $B$ comprising the second and third qubits.  Consider the state
$(1/2)(|00\rangle+|11\rangle) \otimes (|00\rangle+|11\rangle)$.  Then, measure $B$ in a Bell basis.
However, while the wavefunction is non-negative as written down, it is not non-negative if we use a Bell basis for $B$.

We will show that the non-negativity constraint has important consequences.

\subsection{Without Positivity Constraint}
We begin, however, by considering the problem of teleportation in the absence of the non-negativity constraint, for which wavefunctions such teleportation is possible for {\it some} choice of basis on $B$.  Since in this subsection we allow arbitrary bases on $B$ for the measurement, we are in fact allowing arbitrary basis rotations on subsystems $A,B,C$.  Hence, the state (after rotation) can be written as
\be
\label{rotate}
\psi=\psi_{A,B_L} \otimes \psi_{B_R,C},
\ee
where the Hilbert space ${\cal H}_B$ on $B$ is a tensor product of two Hilbert spaces, ${\cal H}_{B_L}\otimes {\cal H}_{B_R}$, and $\psi_{A,B_L},\psi_{B_R,C}$ are wavefunctions on the pairs of Hilbert spaces ${\cal H}_A \otimes {\cal H}_{B_L}$ and ${\cal H}_{B_R}\otimes {\cal H}_C$.  In fact, condition (\ref{fac}) is equivalent to the statement that, up to rotation, condition (\ref{rotate}) above holds.

The two wavefunctions $\psi_{A,B_L},\psi_{B_R,C}$ are fully parametrized, up to rotations on $A,B,C$, by their Schmidt coefficients.  Let $\lambda(1)\geq \lambda(2) \geq ...$ be the sequence of Schmidt coefficients for $\psi_{A,B_L}$ and let $\gamma(1) \geq \gamma(2) \geq ...$ be the sequence of Schmidt coefficients for $\psi_{B_R,C}$.
We will ask, under which conditions is it possible that 
e to achieve the situation that the {\it average} entanglement entropy between $A$ and $C$ {\it after} measurement is equal to the entanglement entropy between $A$ and $BC$ before measurement.  
Let $\rho_A$ be the reduced density matrix on $A$ and let $\rho_A(j)$
be the reduced density matrix on $A$ conditioned on measurement outcome $j$.  Let measurement outcome $j$ occur with probability $P^B_j$.
Since $\rho_A$ is a convex combination of $\rho_A(j)$, by convexity of von Neumann entropy, it follows that the entropy of $\rho_A$ is greater than or equal to the average entropy
of $\rho_A(j)$:
\be
\label{convex}
S(\rho_A)\geq \sum_j P^B_j S(\rho_A(j)),
\ee
where $S(...)$ denotes the von Neumann entropy of a density matrix.
Further, equality can only be attained in the case that
\be
\label{equal}
\rho_A=\rho_A(j)
\ee
for all $j$ such that $P^B_j>0$.
Hence equality can only be achieved when, for every measurement outcome on $B$, the sequence of Schmidt coefficients between $A$ and $C$ of the resulting state is identical with the sequence $\lambda$.
We will assume without loss of generality that $\rho_A$ has no zero eigenvalues.

We look for conditions under which this is possible, allowing $d_B$ to be arbitrarily large (equivalently, we can take $d_B=d_A d_C$ and then allow an arbitrary POVM measurement on $B$ rather than a projective measurement).
We claim that 
\begin{lemma}
For any given $\rho_A$ and $\rho_C$,
there exist states $\psi$ which obey Eq.~(\ref{fac}) and Eq.~(\ref{equal})
if and only if the sequence of eigenvalues of $\rho_C$ is majorized by the sequence $1/d_A,1/d_A,....,1/d_A,0,...$, where there are $d_A$ non-zero entries and the remaining $d_C-d_A$ entries are zero.
\begin{proof}
Let us denote this sequence $1/d_A,...$ as $\tau$.
Here, we say that one sequence $a$ majorizes another sequence $b$ if, when the entries are sorted in descending order so that the sequences are $a_1 \geq a_2 \geq a_3 \geq ...$ and $b_1 \geq b_2 \geq b_3 \geq ...$ we have $\sum_{k=1}^l a_k \geq \sum_{k=1}^l b_k$ for all $l$.

We first show this claim in the case that $\rho_A$ is maximally mixed so that all eigenvalues are equal to $1/d_A$.

Consider first the direction of showing that this condition is necessary.  Let $\rho_C(j)$ denote the reduced density matrix on $C$ after measurement outcome $j$, so that $\rho_C=\sum_j P^B_j \rho_C(j)$.  We have that all $\rho_C(j)$ have the same eigenspectrum (namely, the sequence $\tau$), and hence $\rho_C(j)=U_j \rho_C(j=1) U_j^\dagger$, for some unitary matrices $U_j$, so $\rho_C=\sum_j P_B(j) U_j \rho_C(j=1) U_j^\dagger$.  Since the sequence eigenvalues of $\rho_C(j)$ is equal to $\tau$, by Uhlmann's theorem\cite{uhlmann}, this condition on $\rho_C$ implies that
the sequence $\tau$ majorizes the sequence of eigenvalues of $\rho_C$, as claimed.

Now we show that  this condition is sufficient.  Again by Uhlmann's theorem, if the sequence $\tau$ majorizes the sequence of eigenvalues of $\rho_C$, then we can write $\rho_C=\sum_v P(v) \rho_C(v)$ for {\it some} sequence $P(v)$ which defines a probability distribution, and some sequence of matrices $\rho_C(v)$ which all have eigenspectrum $\tau$.  Let this sequence have $j$ take one of $K$ possible values for some $K$.
We now given a specific state $\psi$ which realizes this set of $\rho_C(v)$ while still obeying condition (\ref{fac}).  We pick $d_B=d_A^2 K$.
We label basis states on $B$ by a triple $(t,u,v)$, with $0 \leq t \leq d_A-1, 0 \leq u \leq d_A-1$, and $1 \leq v \leq K$.
For each $v$, let $\rho_{AC}(v)$ be a maximally entangled state between $A,C$ such that the reduced density matrix on $C$ is $\rho_C(v)$.
We pick
\be
\psi=\sum_{t,u,v} \frac{1}{d_A} \sqrt{P(v)} Z_A^{t} X_A^{u} |(t,u,v)\rangle_B \otimes \rho_{AC}(v),
\ee
where $Z_A$ is a diagonal unitary on $A$ with diagonal entries $1,\exp(2 \pi/d_A),\exp(4 \pi/d_A),...$ and $X_A$ is the ``shift" operator on $A$: it is a unitary matrix with $0$ entries everywhere except for entries $1$ on the superdiagonal and in the bottom left corner.
To verify that this obeys Eq.~(\ref{fac}), it suffices to check that for each $v$, the state 
$\psi=\sum_{t,u} \frac{1}{d_A} Z_A^{t} X_A^{u} |(t,u,v)\rangle_B \otimes \rho_{AC}(v)$ obeys this condition; this can be verified directly.

We now consider the general case in which the state $\rho_A$ is {\it not} maximally mixed.  We will proceed by reducing this to the previous case.  If $\rho_A$ has some number of zero eigenvalues, we restrict to the Hilbert space orthogonal to those zero eigenvalues, so we can assume that $\rho_A$ has no zero eigenvalues.
Define the state $\tilde \psi$ by
\be
\tilde \psi= Z^{-1/2} \rho_A^{-1/2} \psi,
\ee
where $Z$ is a normalization chosen so that $|\tilde \psi|^2=1$.
This define a new state $\tilde \psi$ which obeys the factorization condition and on which the reduced density matrix $\tilde \rho_A$ is maximally mixed.  Further, let $\tilde \rho_A(j)$ denote the reduced density matrix on $A$ for state $\tilde \psi$ after measurement outcome $j$.  We have, from Eq.~(\ref{equal}), that $\tilde \rho_A=\tilde \rho_A(j)$.  Hence, by the above results applied to state $\tilde \psi$, we find that
the density matrix $\rho_C$ is majorized by $\tau$.
\end{proof}
\end{lemma}

Finally, we remark that this calculation we have done amounts to a bound on the so-called ``entanglement of assistance"\cite{eofa} for a particular class of mixed states, $\rho_{AC}=\rho_A \otimes \rho_C$.  As shown in Ref.~\onlinecite{eofaa}, the asymptotic entanglement of assistance for such a state (asymptotic in the sense of given many copies of the state) is equal to ${\rm min}(S(\rho_A),S(\rho_C))$, while here we see that for a single copy the bound can be smaller.

\subsection{With Positivity Constraint For $A$ Being A Qubit}
We now consider the case under which the non-negativity constraint is imposed.
We first investigate the same question as above: when is it possible to achieve the situation that the {\it average} entanglement entropy between $A$ and $C$ {\it after} measurement is equal to the entanglement entropy between $A$ and $BC$ before measurement? So, as above we may assume that Eq.~(\ref{equal}) holds.  Without loss of generality we assume indeed that all $P^B_j>0$.  We again allow $d_B$ to be arbitrarily large.

We begin with some generalities before specializing to the case $d_A=2$ in the lemma at the end.  
For given outcome $j$, the resulting normalized wavefunction is equal to
\be
\sum_{i,k} \frac{1}{\sqrt{P^B_j}} \psi(i,j,k) |i\rangle \otimes |k\rangle=\sum_i |i \rangle \otimes |v_i(j)\rangle,
\ee
where we define
\be
|v_i(j)\rangle=\sum_k \frac{1}{\sqrt{P^B_j}} \psi(i,j,k) |k\rangle.
\ee
Hence, the condition Eq.~(\ref{equal}) is equivalent to
\be
\label{innerprod}
\langle v_l(j) | v_m(j) \rangle = (\rho_A)_{lm},
\ee
for all $j$,
where the right-hand side denotes the $l,m$ matrix element of $\rho_A$.

From the factorization equation Eq.~(\ref{fac}) we have that
\be
\label{outer}
\sum_j P^B_j |v_l(j) \rangle \langle v_m(j)| = (\rho_A)_{lm} \rho_C.
\ee

For a vector such as $v_l(j)$, let $|v_l(j)|_1$ denote the $1$-norm of the vector, namely the sum of the absolute values of the entries of the vector.
This is equal to $\sum_k \psi(l,j,k)$.
For a matrix $\rho$, let $|\vec \rho|_{1}$ denote the sum of the absolute values of the entries of the matrix.  We use superscript $\vec \ldots$ to indicate that we regard the matrix as a vector and then apply the $1$-norm for that vector, to distinguish this from the usual $1$-norm of a matrix.

We will use this $|\vec \ldots |_1$ norm frequently, so let us pause to motivate it.  Suppose we have some non-negative wavefunction $\phi$ on $ABC$, with given $d_C$.
Let $\sigma_{AC}$ denote the reduced density matrix of $\phi$ on $AC$.  Suppose that $\sigma_{AC}$ does not factorize as $\sigma_A \otimes \sigma_C$.  Let us try to construct
a state $\psi$ that will obey the factorization condition.  The state $\psi$ will have a Hilbert space on $B$ that is the tensor product of the Hilbert space of $\phi$ on $B$ with a new degree of freedom taking $d_C!$ possible values.
Let $\psi$ have amplitude
\be
\label{psifromphi}
\psi(i,(j,\pi),k)=\frac{1}{\sqrt{d_C!}}\phi(i,j,\pi(k)),
\ee
where here the notation $(j,\pi)$ denotes a state on $B$ labelled by a pair $j=1,2$ and $\pi$ defining a permutation, and $\pi(k)$ is the permutation applied to the element $\pi$.
Then, the reduced density matrix $\rho_{AC}$ of $\psi$ is given by $\rho_{AC}=\frac{1}{d_C!} \sum_{\pi} \pi_C \sigma_{AC} \pi_C^{-1}$, where $\pi_C$ is the operator corresponding to the permutation $\pi$ applied to $C$.  This can be understood as applying a quantum channel to $\sigma_{AC}$ to obtain $\rho_{AC}$.  This quantum channel has two invariant subspaces.  One is the one-dimensional space parallel to the vector $(1,1,...)$ with all entries equal; the other is the $d_C-1$ space perpendicular to that.  The quantum channel maximally mixes in each subspace.  Hence, to verify Eq.~(\ref{fac}), it suffices to show that 
\be
\label{verify}
{\rm tr}_C(\rho_{AC} \Pi_{(1,1,...)}) \propto \rho_A,
\ee
where $\Pi_{(1,1,....)}$ projects onto the space parallel to the vector $(1,1,...)$.  However,
the trace of a density matrix with $\Pi_{(1,1,...)}$ is simply proportional to the $|\vec \ldots |_1$ norm of that density matrix, motivating our study of this object.
This construction can be understood as follows: we have added some additional degree of freedom to $B$ (taking $d_C!$ possible values) and used this degree of freedom to apply a unitary transformation to $C$, thus trying to make $C$ more mixed to help satisfy Eq.~(\ref{fac}).  However, we choose these unitaries to be positive (and hence permutations) in order to maintain positivity.

From Eq.~(\ref{outer}) we find that
\be
\label{find}
\sum_j P^B_j |v_l(j)|_1 |v_m(j)|_1 = (\rho_A)_{lm} |\vec \rho_C|_{1}.
\ee
It is in this step that we used the non-negativity assumption: for any given $j$ the $|\vec\ldots|_1$ norm of  $|v_l(j) \rangle \langle v_m(j)|$ is equal to $|v_l(j)|_1 |v_m(j)|_1$, and because of the non-negativity assumption, the $|\vec{\ldots}|_1$ norm of $\sum_j P^B_j |v_l(j) \rangle \langle v_m(j)|$ is simply the sum of the  norms for each $j$.

Without loss of generality, we may assume that $|v_l(j)|=|w(j)|=1/\sqrt{d_A}$ for all $j$.  To see this, consider the initial state $\psi(i,j,k)$.  If the probability of having $i$ be in any given state is equal to zero, the problem reduces to a problem with $d_A$ reduced by $1$.  So, we may assume that the probability of being in any given state on $A$ is nonzero.  Then, we consider a new state $\psi'(i,j,k)$ given by $\psi'(i,j,k)=f(i) \psi(i,j,k)$ where $f(i)$ is some function chosen to ensure that $\psi'$ has equal probability of each state on $A$.  Then, Eq.~(\ref{fac}) for $\psi$ is equivalent to Eq.~(\ref{fac}) for $\psi'$ and the condition of equality of entropy of $\psi$ is the same as that for $\psi'$.

For notational simplicity, let $\tilde v_l(j)=\sqrt{d_A} v_l(j)$.  This is just done to normalize all vectors to $1$.  
We have from Eqs.~(\ref{outer},\ref{find}) that
\be
\label{combined}
\sum_j P^B_j  |\tilde v_l(j)|_1 |\tilde v_m(j)|_1 = |\vec \rho_C|_1 \sum_j P^B_j \langle \tilde v_m(j) | \tilde v_l(j) \rangle.
\ee
Note that since Eq.~(\ref{equal}) holds, the right-hand side is simply equal to $|\vec \rho_C|_1 (\rho_A)_{lm}$ as $\langle \tilde v_m(j) | \tilde v_l(j) \rangle$ is independent of $j$.
In the special case that $l=m$, Eq.~(\ref{combined}) gives that
\be
\sum_j P^B_j |\tilde v_l(j)|_1^2=|\vec \rho_C|_1.
\ee
Hence, $\sum_j P^B_j \Bigl( |\tilde v_l(j)|_1^2+|\tilde v_m(j)|_1^2\Bigr)=2|\vec \rho_C|_1$ and so by Eq.~(\ref{combined}),
\be
\label{sat}
2 \frac{\sum_j P^B_j  |\tilde v_l(j)|_1 |\tilde v_m(j)|_1}{\sum_j P^B_j \Bigl( |\tilde v_l(j)|_1^2+|\tilde v_m(j)|_1^2\Bigr)} = \sum_j P^B_j \langle \tilde v_m(j) | \tilde v_l(j) \rangle.
\ee

Now, let us consider for which cases Eq.~(\ref{sat}) can be satisfied.  That is, given a specified value for $d_C$ and for $\langle \tilde v_m(j) | \tilde v_l(j) \rangle$ (recall that this inner product is independent of $j$), is it possible to satisfy Eq.~(\ref{sat})?
The left-hand side of Eq.~(\ref{sat}) is a ratio of two sums.  By adjusting the probabilities $P^B_j$ we will be able to satisfy the equality unless it is the case that for all $j$ we have that
\be
\label{fcase}
{\rm First \, Case:} \quad 2 \frac{|\tilde v_l(j)|_1 |\tilde v_m(j)|_1}{|\tilde v_l(j)|_1^2+|\tilde v_m(j)|_1^2} >  \langle \tilde v_m(j) | \tilde v_l(j) \rangle,
\ee
or it is the case that for all $j$ we have that 
\be
{\rm Second \, Case:} \quad
2 \frac{|\tilde v_l(j)|_1 |\tilde v_m(j)|_1}{|\tilde v_l(j)|_1^2+|\tilde v_m(j)|_1^2} <  \langle \tilde v_m(j) | \tilde v_l(j) \rangle.
\ee

In fact, what we will find is that for sufficiently small $d_C$ it may not be possible to satisfy Eq.~(\ref{sat}),  because in fact for small enough $d_C$ it might not be possible to find two non-negative vectors $v,w$, with $|v|=|w|=1$ such that 
$2 \frac{|v|_1 |w|_1}{|v|_1^2+|w|_1^2} \leq \langle v|w \rangle$, so that we will always be in the first case above (\ref{fcase}).
Note, it is always possible to find two non-negative vectors $v,w$ such that
$2 \frac{|v|_1 |w|_1}{|v|_1^2+|w|_1^2} >  \langle v|w \rangle$ for any $\langle v|w\rangle<1$ and any $d_C \geq 2$ (obviously, if $d_C=1$ then there is no way to find two vectors with inner product less than $1$).  Simply choose $v=(1,0,0,...)$ and $w=(\cos(\theta),\sin(\theta),0,0,...0)$.  A direct calculation shows that indeed the inequality holds.

In general, given a dimension $d_C$ and a value $\langle v|w|\rangle$, we say that ``the pair $d_C,\langle v|w\rangle$ is {\it feasible}" is
 it is possible to achieve $2 \frac{|v|_1 |w|_1}{|v|_1^2+|w|_1^2} \leq  \langle v|w \rangle$ for two vectors $v,w$ with non-negative entries.
In the appendix, we consider which pairs are feasible.  We find that for $d_C=2$, it is not possible; however, for $d_C>2$, it is possible to achieve this, so long as $\langle v | w \rangle$ is not too small.  The possible values of $\langle v | w \rangle$ tend to zero as $d_C$ tends to infinity.

We finally restrict to qubits and show that the ability to achieve this inequality $2 \frac{|v|_1 |w|_1}{|v|_1^2+|w|_1^2} \leq  \langle v|w \rangle$ is the only restriction, if $d_B$ is allowed to be large enough.
That is,
\begin{lemma}
Let $\rho_A$ be a density matrix with non-negative entries for $A$ being a qubit.
Suppose that $\rho_A$ has equal probability to be in the up or down state.  Let $X_A$ denote the Pauli $X$ operator on $A$.  For a given $d_C$, there exist a $d_B$ and
a non-negative wavefunction $\psi$ on $ABC$ obeying Eqs.~(\ref{fac},\ref{equal}) if and only if
the pair $d_C,{\rm tr}(\rho_A X_A)$ is feasible.

More generally, if $\rho_A$ has probabilities $P_\uparrow,P_\downarrow \neq 0$ of being in the up or down state.  For a given $d_C$, there exist a $d_B$ and
a non-negative wavefunction $\psi$ on $ABC$ obeying Eqs.~(\ref{fac},\ref{equal}) if and only if
the pair
$$d_C, \frac{{\rm tr}(\rho_A X_A)}{2\sqrt{P_{\uparrow} P_{\downarrow}}}$$ is feasible.
\begin{proof}
The general case can be reduced to the first case by considering a state $\psi'$ with amplitudes $\psi'(i,j,k)=(2P^A_i)^{-1/2}\psi(i,j,k)$, where $P^A_i$ is the probability of being in state $i$ on system $A$.

Let $v(j)=v_{\uparrow}(j)$ and $w(j)=v_\downarrow(j)$.
Now, if Eq.~(\ref{equal}) holds, then $\langle v(j) | w(j) \rangle = {\rm tr}(\rho_A X_A)$.
If the pair $d_C,{\rm tr}(\rho_A X_A)$ is not feasible, then Eq.~(\ref{fcase}) holds for all $j$, and so Eq.~(\ref{sat}) cannot be satisfied.  That shows the ``only if" direction.
Now we show the ``if" direction.
Suppose the pair $d_C,{\rm tr}(\rho_A X_A)$ is feasible.  Then, we can choose two pairs of vectors $v(1),w(1)$ and $v(2),w(2)$, with all vectors normalized to unity and $\langle v_1 | w_1 \rangle = \langle v_2 | w_2 \rangle = {\rm tr}(\rho_A X_A)$
such that
the pair $v(1),w(1)$ has
\be
2 \frac{| v(1)|_1 | w(1)|_1}{| v(1)|_1^2+| w(1)|_1^2} \geq  \langle  v(1) |  w(1) \rangle
\ee
and
\be
2 \frac{| v(2)|_1 | w(2)|_1}{| v(2)|_1^2+| w(2)|_1^2} \geq  \langle  v(2) |  w(2) \rangle
\ee
 and find probabilities $P^B(1),P^B(2)$ such that
\be
2 \frac{\sum_j P^B_j |\tilde v(j)|_1 |\tilde w(j)|_1}{|\sum_j P^B_j \Bigl( \tilde v(j)|_1^2+|\tilde w(j)|_1^2\Bigr)} = \sum_j P^B_j \langle \tilde v(j) | \tilde w(j) \rangle.
\ee
So, these probabilities can be used to a give a state with $d_B=2$ that obeys Eq.~(\ref{sat}) and Eq.~(\ref{equal}).  Call this state $\phi$.

However, such a state $\phi$ will not, in general, obey Eq.~\ref{fac}).  As above, define a state $\psi$ which has $d_B=2\cdot d_C!$, using Eq.~(\ref{psifromphi}).  Then, our calculations with Eq.~(\ref{sat}) simply amount to verifying Eq.~(\ref{verify}), as
the trace of a density matrix with $\Pi_{(1,1,...)}$ is simply proportional to the $|\vec \ldots |_1$ norm of that density matrix.
\end{proof}
\end{lemma}

\subsection{Qudits, Entanglement Decrease}
We now consider the qudit case.  In this section, we show that
\begin{prop}
\label{PropDecrease}
Consider any finite $d_A,d_C$.  Then there is a constant $c<1$ such that the following holds.
Assume Eq.~(\ref{fac}) holds.  Assume $\psi(i,j,k)$ is non-negative.
Then,
\be
\label{leqc}
\sum_j P^B_j S(\rho_A(j)) \leq c \cdot {\rm max}(S(\rho_A),S(\rho_C).
\ee

Further,
\be
\label{leqcavg}
\sum_j P^B_j S(\rho_A(j)) \leq \tilde c \cdot \frac{S(\rho_A)+S(\rho_C)}{2},
\ee
for $\tilde c=(1+c)/2$.
\end{prop}
We conjecture that this result can be improved to show that the constant $c$ can be chosen independently of $d_A,d_C$.  In the discussion, we explain why this stronger result would be more useful.

Note that Eq.~(\ref{leqcavg}) follows as an almost immediate corollary of Eq.~(\ref{leqc}):
by convexity, $\sum_j P^B_j S(\rho_A(j)) \leq S(\rho_A)$.  Similarly, since $S(\rho_A(j))=S(\rho_C(j))$,
we have $\sum_j P^B_j S(\rho_A(j)) \leq S(\rho_C)$, so $\sum_j P^B_j S(\rho_A(j)) \leq {\rm min}(S(\rho_A),S(\rho_C))$.
Using this last inequality with Eq.~(\ref{leqc}), Eq.~(\ref{leqcavg}) follows after some algebra.

To prove this proposition,
we first show a simple result that replaces the inequality $\leq$ in Eq.~(\ref{convex}) with a strict $<$ inequality:
\begin{lemma}
\label{lessthanlemma}
Consider any finite $d_A,d_C$.  Assume Eq.~(\ref{fac}) holds.  Assume $\psi(i,j,k)$ is non-negative.
Suppose that $S(\rho_A),S(\rho_C)>0$.  Then,
\be
\label{lessthan}
\sum_j P^B_j S(\rho_A(j)) < {\rm max}(S(\rho_A),S(\rho_C).
\ee
\begin{proof}
Suppose not.  Let $\psi$ be a density matrix such that equality holds with $S(\rho_A)={\rm max}(S(\rho_A),S(\rho_C))$.  Then, we must have that $S(\rho_A)=S(\rho_C)$.
We first show that $\rho_A$ and $\rho_C$ must both have the property that they are proportional to projectors and that they have the same rank; that is, there is some $\lambda>0$ such that all eigenvalues of $\rho_A$ and $\rho_C$ are either equal to $\lambda$, or are equal to $0$.  To see this, note that equality $S(\rho_A)={\rm max}(S(\rho_A),S(\rho_C))$ can only hold if $\rho_A(j)$ is independent of $j$ and equal to $\rho_A$ and similarly, $\rho_C(j)$ must also be independent of $j$ and equal to $\rho_C$.  Suppose $\rho_A$ has eigenvalues $\lambda_1 \geq \lambda_2 \geq ...$, with corresponding eigenvectors $v_1,v_2,...$   The matrix $\rho_C$ must have the same eigenvalues $\lambda_1,\lambda_2,...$ 
Then, in the eigenbasis of $\rho_A$ and eigenbasis of $\rho_C$, 
\be
\psi(j)=\sum_{\alpha} \sqrt{\lambda_\alpha} U(j)_{\alpha\beta} |\alpha\rangle_A \otimes |\beta\rangle_C,
\ee
where $\psi(j)$ denotes the pure state on $AC$ conditioned on measurement outcome $j$ and
where $|\alpha\rangle_A,|\beta\rangle_C$ in this equation denote basis vectors in the eigenbasis of $\rho_A,\rho_C$.  The matrix $U(j)$ is a unitary matrix which commutes with $\rho_A$; that is, $U_A$ is block diagonal in this eigenbasis, only mixing basis vectors with the same eigenvalue.
Suppose that $\rho_A$ is not a projector.  Choose any two distinct nonzero eigenvalues $\lambda_>>\lambda_<>0$.  Let $\Pi_A(x)$ project onto the subspace of eigenvectors of $\rho_A$ with eigenvalue $x$ and let $\Pi_C(x)$ project onto the subspace of eigenvectors of $\rho_C$ with eigenvalue $x$.  Then, we find that $${\rm Tr}\Bigl((\rho_{AC}-\rho_A\otimes\rho_C) \Pi_A(\lambda_>) \Pi_C(\lambda_<)\Bigr) \neq 0,$$ contradicting the assumption Eq.~(\ref{fac}); to see that this is nonzero, note that it has the same nonzero value for any choice of $U$.

So, indeed $\rho_A,\rho_C$ must both be proportional to projectors and must have the same rank.
Recall that $\rho_A$ has non-negative entries.  We claim that a projector with non-negative entries must be block diagonal, and in each block must either be zero or have a single nonzero eigenvector; this eigenvector is a non-negative vector $v$ and the projector in that block is the outer product of $v$ with itself.
That is, such a projector $P$ can be written as
\be
\label{Pform}
P=\begin{pmatrix}
P_{1} & \\
& P_{2} \\
&& P_{3} \\
&&&...
\end{pmatrix},
\ee
where each $P_{ii}$ is a block matrix which is either equal to zero, or has matrix elements $(P_{i})_{ab}=v_a v_b$ for some non-negative vector $v$ with norm $1$.
To see Eq.~(\ref{Pform}), pick any vector $w$ which is $1$ in some coordinate and $0$ in all other coordinates.  Consider $Pw$.  If $Pw$ is zero, then the $w$-th row and column of $P$ is equal to zero, giving us one of the zero blocks of Eq.~(\ref{Pform}).  If $Pw$ is nonzero, then the result is some eigenvector of $P$ with eigenvalue $1$.  Normalize this vector and call it $v$.  Let $C$ be the set of coordinates on which $v$ is nonvanishing.  
We claim that this gives us one of the nonzero blocks of $P$.  To see this, consider any normalized eigenvector of $P$ orthogonal to $v$.  Suppose that this eigenvector $w$ is nonvanishing on some coordinates in $C$.  Then, since $v$ and $w$ are orthogonal, $w$ must be nonzero and positive on some set of coordinates $C_+ \subset C$ and nonzero and negative on some other set of  coordinates $C_- \subset C$.  Then, $P$ must have vanishing matrix elements between all coordinates in $C^+$ and coordinates in $C^-$ (if not, by replacing $w$ with a new vector obtained by replacing all entries of $w$ with their absolute value, we would find a vector $w'$ with $(w',P w')>1$, which is a contradiction).
However, this contradicts our assumption on how $C$ was obtained: $P$ has a nonvanishing matrix element between the coordinate in which $w$ is nonzero and every other coordinate in $C$ and so $P$ has nonvanishing matrix elements between any pairs of coordinates in $C$.
Here we use a property of projection operators with non-negative entries: if matrix element $P_{ab}$ and $P_{bc}>0$, then $P_{ac}>0$; this follows from the fact that $P^2=P$.

So, suppose that $(1/\lambda) \rho_A$ has the form (\ref{Pform}).  Let $|\alpha\rangle_A$ denote basis vectors for states on $A$ such that
\be
|\alpha\rangle_A\langle \alpha| =
\label{Pform2}
\begin{pmatrix}
0 & \\
& 0\\
&& ... \\
&&& P_{\alpha} \\
&&&& 0 \\
&&&&& ...
\end{pmatrix}
\ee
Let $|\alpha\rangle_C$ be similar states on $C$ (note that the matrices $P_{\alpha}$ will mean different things depending on whether we are referring to $A$ or $C$).
Then, $\psi=\sum_{\alpha,j,\beta} \psi(\alpha,j,\beta) |\alpha\rangle_A \otimes |j\rangle_B \otimes |\beta\rangle_C$ for some $\psi(\alpha,j,\beta)$, where in an abuse of notation
$\psi(\alpha,j,\beta)$ refers to a possibly different function from the original $\psi(i,j,k)$.  Further, $\psi(\alpha,j,\beta)$ is still non-negative.
Both $\rho_A$ and $\rho_C$ are proportional to the identity matrix in this basis, as each block of $\rho_A,\rho_C$ in the original basis becomes
a single diagonal entry.  By Eq.~(\ref{fac}), $\rho_A \otimes \rho_C$ is also proportional to the identity matrix in this basis.
Hence, by the non-negativity of $\psi(\alpha,j,\beta)$, it follows that if $\psi(\alpha,j,\beta)>0$ for some specific choice of $\alpha,j,\beta$ then $\psi(\alpha',j,\beta')=0$ unless $\alpha=\alpha'$ and $\beta=\beta'$, as if this were not true then $\rho_A \otimes \rho_C$ would have
non-vanishing off-diagonal entries.  Hence, after the measurement of $j$, the states $A$ and $C$ are completely {\it unenentangled}, rather than maximally entangled.
\end{proof}
\end{lemma}

We now show a result on how the entropy behaves in the case that $\rho_A,\rho_C$ each have one large eigenvalue and all other eigenvalues small.  Note that this next lemma does not make use of the nonnegativity assumption.
\begin{lemma}
\label{OneBig}
For any finite $d_A,d_C$, for any $c<1$, there is an $\epsilon>0$ such that the following holds.
Assume Eq.~(\ref{fac}) holds.
Suppose that both $\rho_A,\rho_C$ have the property that all but one of their eigenvalues are less than or equal to $\epsilon$.
Then,
\be
\label{leqcweak}
\sum_j P^B_j S(\rho_A(j)) \leq c \cdot {\rm max}(S(\rho_A),S(\rho_C)).
\ee
\begin{proof}
Let the second largest eigenvalue of $\rho_A$ be $\lambda_2(A)$ and let the second largest eigenvalue of $\rho_C$ be $\lambda_2(C)$.  Assume without loss of
generality that $\lambda_2(A)\geq \lambda_2(C)$.  Then, ${\rm max}(S(\rho_A),S(\rho_C)) \geq -\lambda_2(A) \log(\lambda_2(A))$.
Let $\psi(j)$ be the pure state on $AC$ conditioned on outcome $j$.  Let $\Pi_{small,A}$ denote the projector on $A$ orthogonal to the largest eigenvalue of $\rho_A$ and let $\Pi_{small,C}$ denote the projector on $C$ orthogonal to the largest eigenvalue of $\rho_C$.

For any given $j$, let $P_{small}(j)$ be given by
\be
P_{small}(j)=\langle \psi(j)| \Pi_{small,A} \otimes (1-\Pi_{small,C})+(1-\Pi_{small,A})\otimes \Pi_{small,C} |\psi(j)\rangle
\ee
and let $P_{small,small}(j)$ be given by
\be
P_{small,small}(j)=\langle \psi(j)| \Pi_{small,A}  \Pi_{small,C} |\psi(j)\rangle.
\ee
That is, $P_{small,small}(j)$ is the probability that both $A,C$ are in the ``small probability" space (the space orthogonal to the largest eigenvalue) while $P_{small}(j)$ is the probability that
one of them is in the small probability space.
Note that
\be
\sum_j P^B_j P_{small}(j) =\langle \psi| \Pi_{small,A} \otimes (1-\Pi_{small,C})+(1-\Pi_{small,A})\otimes \Pi_{small,C} |\psi\rangle \leq (d_A-1) \lambda_2(A)+(d_C-1)\lambda_2(C),
\ee
and, using Eq.~(\ref{fac}),
\be
\label{b2}
\sum_j P^B_j P_{small,small}(j) \leq (d_A-1)(d_C-1) \lambda_2(A) \lambda_2(C).
\ee

Consider state $\psi(j)$.  
Let $|0\rangle_A,|0\rangle_C$ denote a normalized eigenvector of $\rho_A,\rho_C$ respectively with largest eigenvalue.
We write the state $\psi(j)$ as
$a |0\rangle_A \otimes |0\rangle_C + b |0\rangle_A \otimes |v\rangle_C +c |w\rangle_A \otimes |0\rangle_C+
\chi_{AC}$ where $|v\rangle_C$ and $|w\rangle_A$ are both normalized vectors in the range of $\Pi_{small,C}$ and $\Pi_{small,A}$, respectively, and $\chi_{AC}$ is an unnormalized state in the range of $\Pi_{small,A} \Pi_{small,C}$.
Then, $|b|^2+|c|^2=P_{small}(j)$ and $|\chi|^2=P_{small,small}(j)$.
Let $P_{small,A}(j)=|c|^2$ and $P_{small,C}(j)=|b|^2$.
After some algebra, one finds that ${\rm tr}(\rho_A(j)^2)\geq 1-O(|b|^2 |c|^2 +|\chi|^2)=1-O(P_{small,A}(j) P_{small,C}(j)+P_{small,small}(j))$.  This gives a bound on the second largest eigenvalue of $\rho_A(j)$ and so,
for any fixed dimensions $d_A,d_C$,
we have
\be
\label{fannes}
S(\rho_A(j)) \leq O(-P_{small,A}(j) P_{small,C}(j) \log(P_{small,A}(j) P_{small,C}(j)) - P_{small,small}(j) \log(P_{small,small}(j)).
\ee

We now bound 
$\sum_j P^B_j S(\rho_A(j))$.
Substitute in Eq.~(\ref{fannes}); the term 
$-P_{small,small}(j) \log(P_{small,small}(j))$ is a concave function so the average of this function over $j$ is bounded by the function of the average, i.e., it is bounded by
$-\sum_j P^B_j P_{small,small}(j)  \log(\sum_j P^B_j P_{small,small}(j) )$.
By Eq.~(\ref{b2}), this is bounded by
$O(\lambda_2(A) \lambda_2(C) \log(\lambda_2(A) \lambda_2(C))$, and for sufficiently small 
$\lambda_2(A)$ this is smaller than
$c S(\rho_A)$.
Now consider the term 
$-\sum_j P^B_j
O(P_{small,A}(j) P_{small,C}(j) \log(P_{small,A}(j) P_{small,C}(j))$.
This is bounded by
$-\sum_j P^B_j (P_{small,A}(j)^2 \log(P_{small,A}(j)^2) + P_{small,C}(j)^2 \log(P_{small,C}(j)^2)).$
On the interval $[0,1]$, the function $-x^2 \log(x)$ is bounded by $O(x)$
so 
$-\sum_j P^B_j (P_{small,A}(j)^2 \log(P_{small,A}(j)^2) + P_{small,C}(j)^2 \log(P_{small,C}(j)^2)\leq
O(\sum_j P^B_j P_{small}(j))=O(\lambda_2(A)+\lambda_2(C))$.
For sufficiently small 
$\lambda_2(A)$, this is smaller than
$c S(\rho_A)$.
\end{proof}
\end{lemma}

Now we can show Proposition \ref{PropDecrease}.
Consider a fixed $d_A,d_C$.  Choose any $c<1$ from lemma \ref{OneBig} and the corresponding $\epsilon$.
Consider the set of pairs of density matrices $\rho_A,\rho_C$ such that the second largest eigenvalue of $\rho_A$ is greater than or equal to $\epsilon$ (where $\epsilon$ is chosen from lemma \ref{OneBig}) or the the second largest eigenvalue of $\rho_C$ is greater than or equal to $\epsilon$.  Let $c_1$ be the maximum over density matrices in this set of
$$
\frac{\sum_j P^B_j S(\rho_A(j))}{{\rm max}(S(\rho_A),S(\rho_C)}.$$
Note that the denominator is bounded away from zero on this set so that $c_1$ is bounded.  Since this set is compact, the maximum is achieved somewhere on the set.  However, this implies that $c_1<1$ by lemma \ref{lessthanlemma}.  Choose the $c$ in Proposition \ref{PropDecrease} to be the maximum of this $c_1$ and the constant $c$ from lemma \ref{OneBig}.

\section{One-Dimensional Chain}
We now consider a one-dimensional chain.  We consider a line of qudits, all with the same Hilbert space dimension $d$.  We consider a non-negative pure state on this line such that $\rho_{AB}$ factors as $\rho_A \otimes \rho_B$ for any two disjoint regions $A,B$ which have at least one qudit in between them.  We refer to this as a system with ``zero correlation length".
We number the qudits by integers, $1,...,L$, where we consider a line of $L$ sites.

First,
we claim that given any pair, $i,j$ with $i<j$, if we measure qudits $i+1,i+2,...,i+j-1$ in the computational basis, then the expected entanglement entropy between $1,...,i$ and $j,...,L$ after measurement is superpolynomially small in $j-i$.  

Let us first show a polynomial decay in $j-i$.
First consider any triple $i,i+1,i+2$.  There is some given entanglement across the cuts $i,i+1$ and across the cuts $i+1,i+2$.  This entanglement entropy is at most $\log(d)$ across each cut.  We wish to estimate the expected entanglement across the cut $i,i+2$ after measuring the spin $i+1$.  To apply the results above, one might first imagine that we would treat $1,...,i$ as system $A$ and spins $i+2,...,L$ as system $C$, while letting $B$ be spin $i$.  However, the results above required that $A$ and $C$ both have bounded Hilbert space dimension, and in this case the dimension of $A,C$ would depend on $L$.  To get around this difficulty, note that if we measure spin $1,...,i-1$ and spins $i+3,...,L$ then regardless of the measurement outcome, the entanglement across the cuts $i,i+1$ and $i+2,i+3$ is unchanged.
Further, the expected entanglement across $i,i+2$ after measuring $i+1$ (i.e., expected over different measurement outcomes on $i+1$) is independent of the measurement outcome on $1,...,i-1$ and $i+3,...,L$.  Indeed, we could simply project onto any state in the computational basis with nonzero amplitude on spins $1,...,i-1$ and $i+3,...,L$ giving a wavefunction on the three remaining spins $i,i+1,i+2$.

Then, we can apply the results above and show that the expected entanglement between $i,i+2$ for this three site system is bounded by $\tilde c \log(d)$, where $\tilde c<1$ is from Proposition \ref{PropDecrease}.
Remark: we could in fact bound the expected entanglement more tightly by $c \log(d)$, but we will be applying this proposition several times in a recursive fashion and it will be useful in later steps of this
process to use the bound in terms of $\tilde c$, as we will see; so, for simplicity of presentation, we use the bound in terms of $\tilde c$ here also.
Hence, if we consider the original wavefunction (without projecting onto a state on spins $1,...,i-1$ and $i+3,...,L$) we find that the expected entanglement across the cut $i,i+2$ is bounded by $\tilde c \log(d)$.
We can similarly bound the entanglement across cut $i+2,i+4$ after measuring spin $i+3$, and so on, measuring all spins $i+k$ for odd $k$.  The result is a new spin chain with half as many spins and with the expected entanglement across each cut bounded by $\tilde c \log(d)$.  We then repeat the process, again measuring half the spins (in this case, the spins that are measured are those that in the original chain were labelled $i+2k$ for odd $k$), so that the expected entanglement across each cut is bounded by $\tilde c^2$.  In general, after $m$ steps of this process, the expected entanglement across any cut is bounded by $\tilde c^m$ and the number of steps $m$ we can do before $i,j$ become neighbors is $\lfloor \log_2(j-i) \rfloor$, so that indeed the expected entanglement is polynomially small in $j-i$.
Remark: here we can see why it is useful to use the bound in Proposition \ref{PropDecrease} involving $\tilde c$ rather than $c$; since this bound involves in this case an average (i.e., a linear function), rather than a maximum (i.e., a nonlinear function), it is simpler to compute the expectation after applying this bound over several steps.

To show superpolynomial decay, note that after a given number $m$ of steps, the expected entanglement across each cut is bounded by $\tilde c^m \log(d)$.  So, for any given cut, the probability that the entanglement is greater than $2\tilde c^m \log(d)$ is at most $1/2$, and these events are independent on different cuts.  Consider three neighboring spins (call them $A,B,C$)  such that the entanglement across cuts $A,B$ and $B,C$ is bounded by $2\tilde c^m \log(d)$  for large $m$; for any $\epsilon>0$, for sufficiently large $m$ this implies that we have bounded the second eigenvalues $\lambda_2(A),\lambda_2(C)$ of the reduced density matrices on $A,C$ by $\epsilon$.  So, we may then apply lemma \ref{OneBig} with any desired constant $c'>0$ (i.e., for any $c'>0$, there is a corresponding $\epsilon$) to bound the expected entanglement across the cut $A,C$ after measuring $B$ by $c' \cdot 2\tilde c^m \log(d)$ and so with probability at least $1/2$, the entanglement across $A,C$ is at most $2c' \cdot 2\tilde c^m \log(d)$.
Thus, we have any desired polynomial decay.  For a precise proof one must consider the event that after measuring $B$, the entanglement entropy is larger than $2c' \cdot 2\tilde c^m \log(d)$; this occurs, as we have shown, with probability at most $1/2$, so on an $\Omega(1)$ fraction of cuts, the entanglement entropy after measurement is reduced by $c'$.  
Since the entanglements across a pair of neighboring cuts are independent, the entanglement can be described by a simple random process: on each cut, there is some distribution of values of entanglement.  Then, after each step, the entanglement across a new cut (for example, the entanglement across the cut between $1,3$ after measuring $2$) is bounded by $c'$ times the maximum entanglement across the cuts and also by the minimum entanglement across the cuts.  Since, as mentioned, the entanglement across the cuts is independent, this gives a closed equation for the entanglement distribution across a cut after a given step in terms of the entanglement distribution across the cuts before that step; solving this equation leads to the superpolynomial decay.
We omit the details.

In this way we find that:
\begin{prop}
\label{OneDimDecrease}
For any one-dimensional line of qudits, labelled $1,...,L$, all with dimension bounded by $d$, for any non-negative wavefunction on this line which has zero correlation length in the sense above, given any pair $i,j$, with $i<j$, the expected entanglement entropy between spins $1,...,i$ and spins
$j,...,L$ after measurement of spins $i+1,...,j-1$ in the computational basis is bounded by
\be
\overline{S(\rho_{1,...,i})} \leq \spo(j-i) \log(d),
\ee
where $\spo(\cdot)$ is a superpolynomially decaying function.
\end{prop}

Let us write the wavefunction for this system as $\Psi(\sigma_1,...,\sigma_L)$ where $\sigma_1,...,\sigma_L$ are values of the the spins in the computational basis.
Let $P(\sigma_1,...,\sigma_L)=|\Psi(\sigma_1,...,\sigma_L)|^2$ denote the classical probability distribution associated to the quantum wavefunction on this system.
Given any two subsystems, $A,B$, the mutual information between the associated classical variables is bounded by that between the quantum variables.  That is, if $A_c$ denotes the classical variable obtained by measuring $A$ in the computational basis and $B_c$ denotes that obtained by measuring $B$ in the computational basis, then $I(A_c;B_c)\leq I(A;B)$.
Applying this to the subsystems $1,...,i$ and $j,...,L$ of the above proposition, we find that the mutual information between $(1,...,i)_c$ and $(j,...,L)_c$ conditioned on $(i+1,...,j-1)_c$ is bounded by $\spo(j-i) \log(d)$, where we write $(a,...,b)_c$ to denote the classical variables obtained by measuring $a,a+1,...,b$ in the computational basis.
Hence,
\be
I\Bigl((1,...,i)_c;j_c | (i+1,...,j-1)_c\Bigr) \leq \spo(j-i) \log(d).
\ee
So, by Pinsker's inequality, the one-norm distance between the probability distribution $P(\sigma_1,...,\sigma_j)$ and the probability distribution $P(\sigma_1,...,\sigma_j-1) P(\sigma_{i+1},...,\sigma_j)/P(\sigma_{i+1},...,{j-1})$ is bounded by
\be
\Vert P(\sigma_1,...,\sigma_j)-P(\sigma_1,...,\sigma_j-1) P(\sigma_{i+1},...,\sigma_j)/P(\sigma_{i+1},...,{j-1})\Vert+1 \leq
\frac{1}{2\sqrt{2}}\sqrt{\spo(j-i) \log(d)}.
\ee
Hence, iterating this we find that
\begin{eqnarray}
\label{close}
&&\Vert P(\sigma_1,...,\sigma_L)-\frac{\prod_{i=1}^{L-l} P(\sigma_i,...,\sigma_{i+l})}{\prod_{i=2}^{L-l} P(\sigma_i,...,\sigma_{i+l-1})}
\Vert_1 \\ \nonumber
 &\leq &\frac{L}{2\sqrt{2}}\sqrt{\spo(l) \log(d)}.
\end{eqnarray}

Let $\Psi_{Gibbs}$ be defined as
\be
\Psi_{Gibbs}(\sigma_1,...,\sigma_L)=\sqrt{\frac{\prod_{i=1}^{L-l} P(\sigma_i,...,\sigma_{i+l})}{\prod_{i=2}^{L-l} P(\sigma_i,...,\sigma_{i+l-1})}}.
\ee
We refer to this as a ``coherent Gibbs state", since it can be written as
\be
\Psi_{Gibbs}(\sigma_1,...,\sigma_L)=\exp\Bigl(\frac{1}{2}\sum\prod_{i=1}^{L-l} \log(P(\sigma_i,...,\sigma_{i+l}))
-\frac{1}{2}\sum_{i=2}^{L-l} \log(P(\sigma_i,...,\sigma_{i+l-1}))\Bigr),
\ee
and each term in the exponential can be regarded as some local classical Hamiltonian (with interaction range at most $l$) and the exponential can then be regarded as a Gibbs state of this Hamiltonian.
We have the general bound for any two mixed states $\rho,\sigma$ that are diagonal in the computational basis that
\be
{\rm tr}(\sqrt{\rho}\sqrt{\sigma})\geq 1-\frac{1}{2} \Vert \rho-\sigma \Vert_1 \Bigr.
\ee
This bound is a special case of a bound in Ref.~\onlinecite{ref40}.

Applying this bound to the mixed states $\rho,\sigma$ corresponding to the classical probability distributions $P(\sigma_1,...,\sigma_L)$ and $\frac{\prod_{i=1}^{L-l} P(\sigma_i,...,\sigma_{i+l})}{\prod_{i=2}^{L-l} P(\sigma_i,...,\sigma_{i+l-1})}$, and using Eq.~(\ref{close}) we get that
\be
\langle \Psi | \Psi_{Gibbs} \rangle \geq \frac{L}{\sqrt{2}}\sqrt{\spo(l) \log(d)}.
\ee

This construction of the coherent Gibbs states may remind the reader of the Hammersley-Clifford theorem, which states that for any classical probability distribution for spins on a graph, with the property that the measurement of all spins neighboring a given spin $i$ will complete factorize the probability distribution between $i$ and the rest of the system, and with the property that the probability distribution is nonzero on all configurations, then the probability distribution is the Gibbs state of a Hamiltonian that is a sum of terms supported on cliques of the given graph.  Thus, it may seem surprising that we do not need to require that the probability distribution is nonzero here, as that is known to be a necessary condition\cite{hamcliffpos} for the Hammersley-Clifford theorem.  The resolution is that we have considered the system on a one-dimensional chain with open boundary conditions, which gives us a natural ordering of the spins, allowing us to iteratively build up the distribution on the first spin, then first two spins, and so on, as done above.  So, the second spin will help ``shield" the first spin from the rest of the chain (in our problem, this shielding is not perfect in contrast to the Hammersley-Clifford result), while if we had considered the problem with periodic boundary conditions, even under the stronger assumptions of Hammersley-Clifford the second spin alone would not shield the first spin from the rest of the chain, but we would also need to measure the $L$-th spin.

\section{Discussion}
We have shown an intrinsic sign problem for commuting Hamiltonians with TQO-2 in the same phase as the double semion model.
Further, we have shown several bounds on possible teleportation in systems with a non-negativity constraint.  We have finally shown that the wavefunction has a coherent Gibbs state form for one-dimensional systems with zero correlation length.  This form is based on showing that given any pairs of qudits $i,j$, if we measure qudits $i+1,i+2,...,i+j-1$ in the computational basis, then $i,j$ are superpolynomially close to decorrelated.
If Proposition \ref{PropDecrease} could be improved to make the constant $c$ independent of $d_A,d_C$, then we would be able to prove a similar result in two dimensions, in this case showing that given any qudit $i$, if we measure all other qudits within radius $R$ of $i$, then $i$ is approximately decorrelated from the rest of the system.  This would then hopefully allow a proof of an approximate Gibbs state form for two dimensional systems with zero correlation length, a step toward the goal mentioned in the introduction of classifying topological order for positive states, although now because the problem is in two dimensions, the proof of the coherent Gibbs state would be much more difficult (in particular, we would not have such a natural ordering of spins as in one dimension; see the discussion above about shielding one spin from the rest of the chain).
Thus, while at present the two main results in this paper (intrinsic sign problem for certain commuting Hamiltonians in two dimensions, and coherent Gibbs states for certain wavefunctions in one dimension) are not related to each other, perhaps if there is an extension of the coherent Gibbs state result to two dimensions then this will lead to a connection between these results and to a way of strengthening the first result.

The general study of wavefunctions with this non-negativity constraint may be interesting.  It might, for example, be easier to prove area law results for Hamiltonians if we assume that the ground state wavefunction has this property.  It might be even easier if we assume that every term in the Hamiltonian has the
property that all off-diagonal matrix elements are non-positive.
In this regard, it is worth pointing out that even if the Hamiltonian has this property, there still need not be a ``generalized area law" in the sense of Ref.~\onlinecite{genarealaw}.  That paper used quantum expanders\cite{qe1,qe2} to construct a Hamiltonian on four sites on a line, with interactions between only neighboring sites.  The dimensions of the middle two sites were fixed, but the dimensions of the end sites were parametrized by a parameter $d$.  The Hamiltonian had uniformly (in $d$) upper bounded norm and lower bounded spectral gap, but had large (in fact, maximal, i.e. $\log(d)$) entanglement between either end site and the rest of the chain.
The Hamiltonian however did not have the property of having no sign problem.  However, it is not difficult to modify the Hamiltonian so that it does not have
a sign problem.  We spend some time motivating the construction before writing the desired Hamiltonian.
A first guess as to how to do this (this guess does not work) is to replace the unitary matrices $U_1,U_2,U_3$ which are used to define the quantum expander in that work by unitary matrices with non-negative entries.  In fact, the only unitary matrices which have non-negative entries are
permutation matrices, so we would be trying to construct a quantum expander defined by a channel of the form
\be
\label{qmap}
{\cal E}(\rho) =\frac{1}{k} \sum_{a=1}^{k} P_a \rho P_a^\dagger,
\ee
where in general we allow $k$ (potentially more than $3$) unitary matrices.  However, such a channel cannot be a quantum expander because it has at least
two linearly independent fixed points.  One such fixed point is the maximally mixed state, which the other fixed point is the matrix $\rho$ with all entries equal to $1/d$; i.e., this second fixed point is the projector onto a vector with all entries equal to $1/\sqrt{d}$.  So, we need to modify this solution.
Before giving the modification, we explicitly write down the Hamiltonian that would result from applying the construction of Ref.~\onlinecite{genarealaw} to
this channel.
The Hamiltonian is a sum $H_L+H_M+H_R$ with the middle two sites having $k$ states, with basis vectors $|i\rangle$, for $i=1,...,k$, and with
\be
\label{HL1}
H_L=1-\frac{1}{k}\sum_{i,i'=1}^k P_i P_{i'}^\dagger \otimes |i \rangle \langle i' | 
\ee
\be
\label{HR1}
H_R=1-\frac{1}{k}\sum_{i,i'=1}^k |i \rangle \langle i' | \otimes (P_i P_{i'}^\dagger)^T  
\ee
\be
\label{HM1}
H_M=\sum_{a=2}^{k} (|1a\rangle-|a1\rangle)(|1a\rangle-|a1\rangle)^\dagger.
\ee

To use previous results from Ref.~\onlinecite{tpe}, we will take $k$ even, and fix $P_a=P_{a+k/2}^\dagger$ for $1 \leq a \leq k/2$.
First, note that for $k \geq 4$, for random choice of the permutations, there is only a two dimensional space of fixed points (spanned by the two states given above), with a gap to the rest of the spectrum of the map ${\cal E}(\rho)$.  This follows from the fact that the classical stochastic map which maps a probability distribution $\vec p$ (the entries of the vector denote the probabilities of the various states) by
\be
\vec p \rightarrow \frac{1}{k} \sum_{a=1}^k P_i \vec p
\ee
forms a $2$-copy tensor product expander\cite{tpe}.  The quantum channel defined by (\ref{qmap}) has exactly the same action on $\rho$ as this $2$-copy tensor product expander does on the joint probability distribution of two degrees of freedom.  The construction of Ref.~\onlinecite{tpe} uses at least four permutation matrices to do this, so using such permutation matrices requires a modification of the construction of Ref.~\onlinecite{genarealaw}) to increase the dimension of the middle sites to $k=4$ from $3$.
Hence, combining these results from Refs.~\onlinecite{genarealaw},\onlinecite{tpe}, the Hamiltonian given by Eqs.~(\ref{HL1},\ref{HR1},\ref{HM1}) has
a spectral gap with a doubly degenerate ground state.  This Hamiltonian is frustration free.

However, this still does not give the desired result as we wish a unique, highly entangled ground state.  Note however that we can find quantum channels, all of whose Krauss matrices have non-negative entries, such that there is a
unique fixed point which is the maximally mixed state, with a gap to the rest of the spectrum.
Define a projector $\Pi$ which projects onto the first $d/2$ coordinates.  
Consider the quantum channel
\be
\label{qmap2}
{\cal E}(\rho) =q\frac{1}{k} \sum_{a=1}^{k} P_a \rho P_a^\dagger+ (1-q) \Pi \rho \Pi +(1-q) (1-\Pi)\rho (1-\Pi),
\ee
for any $q$ with $0<q<1$.
This has the desired property: by adding the second and third terms, the only fixed point is indeed the maximally mixed state, and using the claimed properties of the
spectrum of the tensor product expander it is easy to show that there is a gap to the rest of the spectrum.

However, such a quantum channel, which uses non-unitary matrices, does not exactly fit into the scheme of Ref.~\onlinecite{genarealaw}.
We now write down a Hamiltonian which is motivated by this quantum channel which slightly generalizes this scheme.
The middle sites will have dimension $k+2$, with states labelled $1,2,...,k,A,B$.  The two states $A,B$ in a sense will in a sense correspond to the
two additional Krauss operators.  The Hamiltonian is given by
$H'_L+H'_R+H'_M$ with
\be
H'_L=H_L+\Delta_L,
\ee
where
\be
\Delta_L=\Pi \otimes  (|1 \rangle-|A\rangle)  (|1 \rangle-|A\rangle)^\dagger+
 (1-\Pi) \otimes (|1 \rangle-|B\rangle)  (|1 \rangle-|B\rangle)^\dagger
\ee
and
\be
H'_R=H_R+\Delta_R,
\ee
where
\be
\Delta_R= (|1 \rangle-|A\rangle)  (|1 \rangle-|A\rangle)^\dagger \otimes \Pi+
 (|1 \rangle-|B\rangle)  (|1 \rangle-|B\rangle)^\dagger \otimes (1-\Pi),
\ee
\be
H'_M=H_M+|AB\rangle\langle AB|+|BA\rangle\langle BA|.
\ee
To analyze this Hamiltonian, consider first the Hamiltonian $H'_L+H'_R+H_M$.  Note that this is equal to $H_L+H_R+H_M+\Delta_L+\Delta_R$.
Both $H_L+H_R+H_M$ is a frustration free Hamiltonian with a spectral gap (it is an easy exercise to verify that the addition of the states $|A\rangle$ and $|B\rangle$ increases the number of ground states but does not remove the spectral gap), so it is lower bounded by $c_0 Q_0$ for some constant $c_0>0$ and some projector $Q_0$.  Similarly, $\Delta_L+\Delta_R$ is lower bounded by $c_1 Q_1$ for some $c_1>0$ and some projector $Q_1$. 
Hence, $H_L+H_R+\Delta_L+\Delta_R$ is lower bounded by ${\rm min}(c_0,c_1)(Q_0+Q_1)$.
We now use the following lemma about gaps for sums of projectors:
\begin{lemma}
Let $Q_0,Q_1$ be projectors.  For any positive semi-definite operator $O$, let $\Delta(O)$ denote the smallest non-zero eigenvalue of $O$.
Then,
\be
\Delta(Q_0 + Q_1) \geq \frac{1}{2} \Delta\Bigl( Q_1 + (1-Q_1) Q_0 (1-Q_1) \Bigr)=\Delta\Bigl((1-Q_1) Q_0 (1-Q_1)\Bigr).
\ee
\begin{proof}
\label{Qprojlemma}
Applying Jordan's lemma to the projectors $Q_0,Q_1$, we can bring both projectors to a block diagonal form with blocks of size at most $2$.  We show that this result holds in each block.  Consider a block of size $2$ in which each projector has one eigenvalue equal to zero and one equal to one (the case of a block of size $1$ can be handled straightforwardly).
Pick a basis in this block so that
is $Q_1$ is
\be
\begin{pmatrix}
1 & 0 \\
0 & 0
\end{pmatrix},
\ee
and $Q_0$ being
\be
\begin{pmatrix}
\cos(\theta)^2 & \cos(\theta) \sin(\theta) \\
\cos(\theta) \sin(\theta) & \sin^2(\theta)
\end{pmatrix}.
\ee
A direct calculation gives that the smallest nonzero eigenvalue of $Q_0+Q_1$ in the given block is $1-|\cos(\theta)|$,
while the smallest nonzero eigenvalue of $Q_1+(1-Q_1) Q_0 (1-Q_1)$ in the given block is $\sin^2(\theta)$.
Since $1-|\cos(\theta)|\geq \frac{1}{2} \sin^2(\theta)$ the result follows.

To show this last inequality, note that it is equivalent to $1-\frac{1}{2}\sin^2(\theta) \geq \cos(\theta)=\sqrt{1-\sin^2(\theta)}$.
Squaring both sides gives the desired result.
\end{proof}
\end{lemma}

Hence, using this lemma,
we can restrict to the range of $1-Q_1$, and show a gap for $Q_0$ restricted to this subspace.
The range of $1-Q_1$ can be obtained by starting with a space of states including only states $1,...,k$ on the middle two sites and applying an isometry
$\sum_{i=2}^k I \otimes |i \rangle \langle + (1/\sqrt{2}) \Pi \otimes (|1\rangle + |A\rangle)\langle 1 | + (1/\sqrt{2}) (1-\Pi) \otimes (|1\rangle + |B\rangle)\langle 1|$ to the first
two sites and 
$\sum_{i=2}^k |i \rangle \langle \otimes I + (|1\rangle + |A\rangle)\langle 1 | \otimes \Pi + (|1\rangle + |B\rangle)\langle 1| \otimes (1-\Pi)$.
Call the product of these isometries $V$.  We find that $V^\dagger (H_L+H_M+H_R) V \geq (1/2) (H_L+H_M+H_R)$ where, in an abuse of notation, the $H_L+H_M+H_R$ on the left hand side acts on the space of states with $k+2$ states on each middle site, while the $H_L+H_M+H_R$ on the right hand side acts on the space of states with only $k$ states on each middle site.
Hence, the results of Ref.~\onlinecite{genarealaw} showing a spectral gap carry over to this case, and $H'_L+H'_R+H_M$ has a doubly degenerate ground state with a spectral gap.

Finally, we show that $H'_L+H'_R+H'_M$ has a unique ground state.  Since we know that $H'_L+H'_R+H_M$ has a lower bound on the gap, it is lower bounded by a constant times the projector onto the ground state.  Using the same lemma \ref{Qprojlemma}, it suffices to
consider the ground state subspace of $H'_L+H'_R+H_R$  and show that the operator $|AB\rangle\langle AB|+|BA\rangle\langle BA|$ vanishes on one of the ground states, but not the other.  Indeed, the unique ground state of $H'_L+H'_R+H'_M$ is highly entangled.

\appendix
\section{Feasible Solutions}
We consider for which dimensions $d$ it is possible to have two $d$-dimensional vectors, $v,w$, with non-negative entries, with $|v|_2=|w|_2=1$, and with given
inner product $\langle v | w \rangle$, such that
\be
\label{ineq1}
2 \frac{|v|_1 |w|_1}{|v|_1^2+|w|_1^2}\leq \langle v | w \rangle.
\ee

Let $|1\rangle$ denote the vector with all coefficients equal to $1/\sqrt{d}$.
Write $v$ as a linear combination $v=\cos(\theta) |1\rangle+\sin(\theta) v^\perp$, for some angle $\theta$ and some $v^\perp$ with $\langle v^\perp | 1 \rangle=0$ and $|v^{\perp}|=1$.
Similarly, write $w$ as a linear combination $w=\cos(\phi)|1\rangle+\sin(\phi) w^\perp$, where also $|w^\perp|_2=1$ and $\langle w^\perp | 1 \rangle=0$.
Let $z=\langle v^\perp | w ^\perp \rangle$.
Then, we must satisfy
\be
\label{ineq2}
2 \frac{\cos(\theta) \cos(\phi)}{\cos(\theta)^2+\cos(\phi)^2} \leq \cos(\theta)\cos(\phi)+\sin(\theta) \sin(\phi) z \leq \cos(\theta) \cos(\phi)+\sin(\theta)\sin(\phi),
\ee
where without loss of generality we assume that $\sin(\theta)>0,\sin(\phi)>0$.

Let $\theta=\rho+\delta/2$ and $\phi=\rho-\delta/2$ for some angles $\rho,\delta$.
Then, the left-hand side of Eq.~(\ref{ineq2}) is equal to
\begin{eqnarray}
&&2 \frac{\Bigl( \cos(\rho) \cos(\delta/2)-\sin(\rho) \sin(\delta/2) \Bigr) \Bigl( \cos(\rho) \cos(\delta/2)+\sin(\rho) \sin(\delta/2) \Bigr)}
{\Bigl( \cos(\rho) \cos(\delta/2)-\sin(\rho) \sin(\delta/2) \Bigr)^2+ \Bigl( \cos(\rho) \cos(\delta/2)+\sin(\rho) \sin(\delta/2)\Bigr)^2}
 \\ \nonumber
&=&
\frac{\cos(\rho)^2\cos(\delta/2)^2-\sin(\rho)^2\sin(\delta/2)^2}{\cos(\rho)^2\cos(\delta/2)^2+\sin(\rho)^2\sin(\delta/2)^2}.
\end{eqnarray}
For any fixed $\delta$, this is minimized by taking $\sin(\rho)^2$ as large as possible.  In this case, ``as large as possible" encounters
the constraint that for $v,w$ having non-negative entries, we have
$\cos(\theta)\geq 1/\sqrt{d},\cos(\phi)\geq 1/\sqrt{d}$.
Without loss of generality, assume $0 \leq \phi \leq \theta$.  So, we can assume that $\cos(\theta)=1/\sqrt{d}$ and so $\sin(\theta)=\sqrt{(d-1)/d}$.
We have $\phi=\theta-\delta$.  Hence, $\cos(\phi)=\cos(\delta)/\sqrt{d}+\sin(\delta)\sqrt{(d-1)/d}$.

Taking this, the left-hand side of Eq.~(\ref{ineq2}) is equal to
\be
\label{eq2}
2 \frac{\cos(\delta)+\sqrt{d-1}\sin(\delta)}{1+\cos(\delta)^2+(d-1)\sin(\delta)^2+2\sqrt{d-1}\cos(\delta)\sin(\delta)}=
2 \frac{\cos(\delta)+\sqrt{d-1}\sin(\delta)}{2+(d-2)\sin(\delta)^2+2\sqrt{d-1}\cos(\delta)\sin(\delta)},
\ee
while the right-hand side is at most equal to $\cos(\delta)$.

By a series analysis, one may verify that the inequality
\be
\label{eq3}
2 \frac{\cos(\delta)+\sqrt{d-1}\sin(\delta)}{2+(d-2)\sin(\delta)^2+2\sqrt{d-1}\cos(\delta)\sin(\delta)}\leq \cos(\delta)
\ee
has solutions with the left-hand side less than $1$ for $d>2$; this is done by considering the behavior for small $\delta$ and showing that the left-hand side decreases more rapidly as a function of $\delta$ to order $\delta^2$.  However, a numerical solution shows that for $d=3$, the inequality can only be satisfied if the left-hand side is greater than $0.94\ldots$.  An explicit pair of such vectors $v,w$ can be written in the form $v=(1,0,0)$, $w=(\sqrt{1-2\epsilon^2},\epsilon,\epsilon)$ for appropriate $\epsilon$

For larger $d$, however, the inequality can be satisfied at smaller values of the left-hand side.  For fixed $\delta$, the left-hand side scales as
$2/(\sqrt{d}\sin(\delta))$ for large $d$.  To make this as small as possible, one would like to take $\sin(\delta)$ large, thus the best asymptotics for the left-hand side we could hope to achieve is $2/\sqrt{d}$.  In fact, we can achieve this asymptotics: pick $\sin(\delta)=\sqrt{1-f(d)^2}$ and $\cos(\delta)=f(d)$ for any function $f(d)$ which is asymptotically less than $1$ but asymptotically greater than $2/\sqrt{d}$.

Finally, we note that Eq.~(\ref{ineq2}) has no solution with the left-hand side less than $1$ for $d=2$.
This can be seen by verifying that in this case the inequality (\ref{eq3}) becomes
\be
\frac{cos(\delta)+\sin(\delta)}{1+\cos(\delta)\sin(\delta)}\leq \cos(\delta),
\ee
or equivalently, $\cos(\delta)+\sin(\delta) \leq \cos(\delta)+\cos^2(\delta)\sin(\delta)$, which for $\sin(\delta),\cos(\delta)>0$ can only be satisfied if $\cos(\delta)=1$.

{\it Acknowledgments---} I thank F. Brandao, M. P. A. Fisher, M. Freedman, A. T. B. Hastings, Z. Wang, and J. Yard for useful discussions.  I thank M. Soleimanifar for pointing out an error in the first version; this version corrects the exponential decay in
Prop. \ref{OneDimDecrease} to superpolynomial decay.

\end{document}